\documentclass[
 reprint,
 amsmath,amssymb,
 aps,
prb,
floatfix,
]{revtex4-2}

\usepackage{placeins}
\usepackage{lipsum}
\usepackage{graphicx}
\usepackage{dcolumn}
\usepackage{xcolor}
\usepackage{bm}
\usepackage{hyperref}
\usepackage[normalem]{ulem}
\usepackage{subcaption}
\usepackage{comment}
\usepackage{subcaption}
\definecolor{revised}{RGB}{0,0,180}

\begin{document}

\title{Twisted light drives chiral excitations of \\ interacting electrons in nanostructures with magnetic field} 

\author{F. J. Rodr\'{\i}guez}

\author{L. Quiroga}%
\affiliation{Departamento de F\'{\i}sica, Universidad de Los Andes, A.A.4976, Bogot\'a D.C., Colombia
}%

\author{N. F. Johnson}
\affiliation{
 Physics Department, George Washington University, Washington, DC 20052, U.S.A.
}%

\begin{abstract}

Ultrafast control of confined two-dimensional few-electron systems requires detailed understanding of light--matter interactions at the nanoscale. Twisted light (TL), a special kind of light carrying orbital angular momentum, offers a powerful tool for driving symmetry-resolved transitions in such quantum-confined structures. We consider a simple yet realistic model in which a TL pulse interacts with a nanostructure containing two interacting, charged electronss (e.g. a semiconductor quantum dot containing 2 electrons) under the influence of a perpendicular magnetic field. Since image charge effects are always present in layered nanostructures, we use an effective electron--electron interaction potential of the form \( 1/r^{\eta} \). For \( \eta = 2 \), the system acquires an underlying \( \mathfrak{su}(1,1) \) dynamical symmetry, which allows for analytical solutions and hence provides a clear framework for interpreting selection rules, parity transitions, and angular-momentum--resolved absorption features. We show that the bare Coulomb \( 1/r \) potential also generates similar energy spectra, demonstrating that the twisted-light--driven excitations are remarkably robust against the specific form of the interaction. We analyze the resulting excitation spectrum with emphasis on its chiral properties. We show that TL pulse excitation, unlike conventional dipolar fields, enables direct access to interaction-driven transitions that are otherwise symmetry-forbidden. In particular, we demonstrate that TL breaks the generalized Kohn theorem, making the internal nanostructure energetic spectrum accessible through multiple-quanta orbital excitations. More broadly, our findings establish TL as a sensitive future probe of correlations, symmetry, and magneto-optical dynamics in strongly interacting confined quantum systems, revealing features that remain dark in standard low-frequency (e.g. infrared) light absorption.\\
\end{abstract}
\maketitle

\section{\label{sec:level1} Introduction}

Particle-particle correlations play a crucial role in the quantum physics of confined multi-body systems. In condensed matter physics, for example, few-body problems have recently taken on
more direct relevance due to
rapid advances in the fabrication of nanodevices, e.g. quantum dots (QDs) containing few electrons \cite{hawrylak}. Semiconductor QDs have  in particular attracted attention in recent decades due to their potential as solid-state platforms for controlled emission of entangled photons, for simulating strongly correlated many-body systems \cite{diepen}, and for allowing fault-tolerant quantum computing protocols \cite{harvey}. Recently, coupled QDs have been exploited to realize an artificial Kitaev chain supporting Majorana electrons \cite{dvir}. Furthermore, highly correlated many-body states, such as the Laughlin state, are known to be a cornerstone concept for understanding macroscale phenomena such as the fractional quantum Hall effect \cite{laughlin}. A number of studies have confirmed the feasibility of achieving precursors of such highly correlated many-body states in few-electron QD systems \cite{johnson,reimann}. 

A particularly impressive example is the experimental generation of Laughlin states of two rotating electronic atoms \cite{baur2008,lunt2024,Deric}, in which a high level of versatility and control of the atom-atom interaction strength is provided by tuning the external magnetic field. In addition, the relevance of two-electron pairing above and below the critical temperature in some superconducting materials which reveals robust electron pairing phases even in the absence of superconductivity \cite{cheng}, heightens the importance of understanding few-electron correlations across a wide range of low-dimensional matter systems.

In parallel with these advances in nanostructure matter systems and atomic physics, there have been a number of significant advances in optics in terms of generating pulses of twisted light (TL) or vortex light, which refers to a kind of beam that carries orbital angular momentum in addition to the conventional spin angular momentum \cite{quinteiro2022}. In particular, the TL transverse profile shows strongly space-varying light
fields -- or equivalently, TL displays optical vortices. Recently, due to their
unique structural features, TL pulses have been widely applied in various fields such as
optical communication  \cite{Wang:16}
and quantum teleportation \cite{Luo:letter}. TL is particularly useful in the study of multilevel quantum systems such as hybrid QD - metal nanoparticle \cite{Mahadavi:pra,Mohanad:scir}, where
the spatial structure of the beam can influence the dynamics of single and collective excitations, and facilitate
processes such as quantum state manipulation, entanglement generation, and quantum information transfer \cite{Drori:sci}.

Twisted light (TL) can also efficiently induce and probe notable optical transitions within Landau levels when a magnetic field is present, because of the energy scales linked to effective confinement and Coulomb interactions \cite{jifuksprb2023}. Specifically, the twisted light's structured helical phase adds a degree of freedom to selectively transfer  photon orbital angular momentum to electrons -- and this could be used to probe electron-electron interactions in new ways, revealing new resonances and topological features.

Given the ultrasmall length and time scales in such few-body systems, it seems that ultrashort TL pulses will likely become a major tool for manipulating and controlling their dynamics. A thorough characterization
of this matter-light interaction is of major fundamental
interest, as it is linked to the long-standing question of
how the quantum dynamics of nanodevices can be controlled by entirely new forms of light.
 
This paper makes two contributions. The first contribution is to future experimental work: We establish experimentally accessible signatures of  electron-electron (e.g. electron-electron, e-e) correlations in a nanostructure with a parabolic confining potential, by leveraging the vortex structure of TL beams. In the rest of the paper, we will for simplicity refer to the typical experimental example where the electrons are electrons and the nanostructures are QDs, though we stress that our results are in principle applicable to a wider class of nanoscale systems than just these. The second contribution of our paper is to future theoretical work: Our paper establishes a new platform for future calculations in such systems by showing that the many-body spectrum, and hence its observable signatures, are remarkably insensitive to the precise form of the electron-electron interaction -- and specifically that an inverse-square form can be used for systems where the interaction is pure Coulomb. This is important since, as we discuss, analytic results can be obtained for an inverse-square (so-called Calogero) interaction. This means  that analytic results can be pursued to obtain other empirical signatures beyond those presented here -- and in particular, the inverse-square form can be reliably used as the basis of perturbation calculations as discussed below.

We note that in most  calculations for quantum dot systems, the e-e interaction is assumed to be a bare Coulomb repulsive potential.
A complete description of this few-electron system is complicated since the confinement
energy, the e-e repulsion, and the cyclotron energy due to applied magnetic fields are typically comparable in magnitude. In a few special situations, analytical results can be found \cite{Taut}. Analytic simplifications of the
exact $N$-particle Hamiltonian or exact solutions of model
$N$-particle Hamiltonians can therefore be useful.
In this work we assume an inverse power-law e-e potential, $1/r^\eta$. In fact, this special form of interaction is
quite realistic in QDs due to the presence of
image charges. For $\eta=2$ (which is referred to as the Calogero model) the interaction resembles
the dipole-like form introduced in \cite{qaj} to allow for
image charge effects: it was shown to successfully  reproduce the angular momentum
ground state transitions and excitation spectrum found
for the Coulomb interaction.
These favorable quantitative comparisons have allowed the exploration of new physical features in multi-electron QDs \cite{jq1995,gonzalez1996}, Quantum Hall systems \cite{jq1997}, nuclear-electron quantum logic gates \cite{rqj2000} as well as cavity control of QD Bell states \cite{brqj2023}.

Specifically, we here extend the analysis of a QD containing interacting electrons in a magnetic field, to now include the effect of twisted light. We show that a dynamical $su(1,1)$ Lie algebra structure can be identified: this then allows us to exploit the integrability of the model  \cite{hakobianintegrability} by using its eigenstates as a basis for numerically solving the system's dynamics under pulsed TL excitation for the general interaction case $1/r^\eta$ with $1 \le \eta \le 2$. A significant TL absorption enhancement is found to come from the interacting electrons' contribution. This is a striking feature that goes significantly beyond earlier work \cite{Hawrylak:prl}, which had by stark contrast found that electron correlation signatures are undetectable for interacting electrons in harmonically confined QDs under dipole coupling to non-vortex light -- which is the so-called Kohn theorem \cite{Brey}.

The paper is organized as follows. Section \ref{sec:level2} introduces the theoretical model and the main features of the e-e interaction model and its close relationship with the Calogero model with an applied magnetic field.  In Section \ref{sec:level3} the interaction between twisted light and a multi-electron QD is addressed and its consequences for an observable such as the number of chiral excitations are analysed. 
Numerical results are presented in Section \ref{sec:level3a}. Conclusions and outlook are summarized in Section \ref{sec:level8}. The Appendices contain a detailed theoretical account of the Lie algebra structure for \( \eta=2 \), as well as a comparative analysis of the effects of the Coulomb ($\eta=1$) and Calogero ($\eta=2$) interactions.

\FloatBarrier
\begin{figure}[htbp]
\centering
\includegraphics[width=1.2\linewidth]{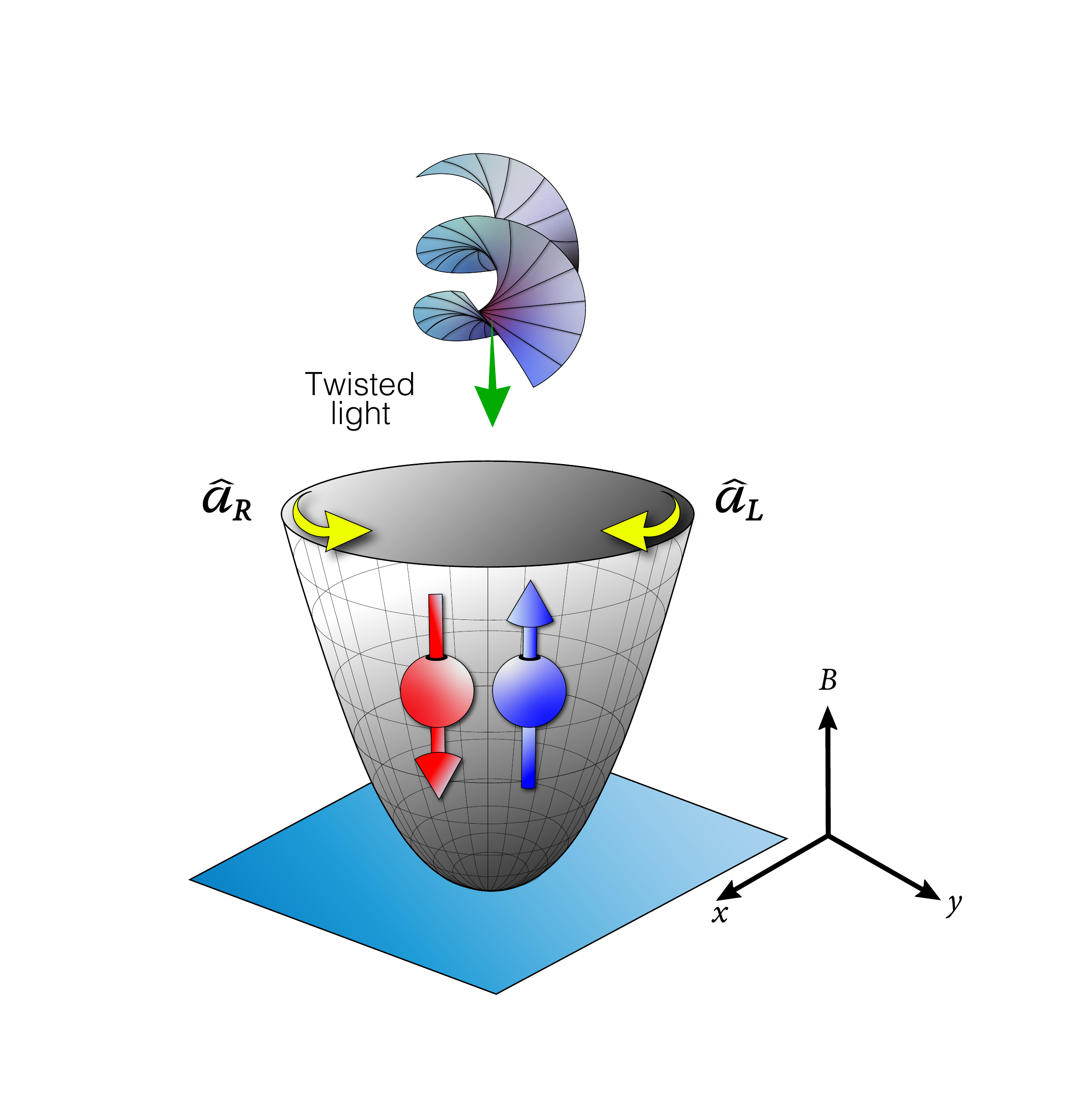}
\caption{\label{fig:fig1} Schematic representation of the system. A twisted-light (TL) pulse is incident on two interacting charged electrons (e.g. electrons) confined in a two-dimensional harmonic quantum dot, under a perpendicular magnetic field \( B \). The chiral operators \( \hat{a}_r \) and \( \hat{a}_l \) destroy right- and left-circular orbital quanta of the relative motion, respectively.}
\end{figure}

\section{\label{sec:level2}Theoretical Model}
\subsection{\label{sec:level2a}Two interacting electrons in a nanostructure }

Figure 1 shows schematically the example system that we use in this paper. It consists of a two-dimensional QD with a harmonic (i.e. parabolic) confining potential that confines two interacting particles (e.g. electrons) under the action of an applied perpendicular magnetic field $B\hat{z}$. 
The model Hamiltonian is:
\begin{eqnarray}
\nonumber \hat{H}_{2e}=\sum_{k=1}^{2} \biggl [  \frac{\hat{{\Pi}}_k^{2}}{2m^*}+\frac{1}{2}m^*\omega_0^2\hat{{r}}_k^{\, 2}+\hat{H}_k^{Z} \biggr ] +V_{e-e}\left ( |\hat{\vec{r}}_1-\hat{\vec{r}}_2| \right )\\
\label{Eq:q1}
\end{eqnarray}
The quantum of energy of the harmonic QD is denoted as $\hbar\omega_0$ and $m^{*}$ is the electron effective mass.

From now on we adopt a homogeneous classical magnetic field of strength $B$ pointing perpendicular to the QD plane, with corresponding vector potential in the symmetric gauge $\hat{{\cal A}}(\vec{r})=\frac{B}{2}(-y,x,0)$. The effects of the magnetic field are twofold: (i) a coupling to the electron orbital motion that comes from the kinetic momentum $\hat{{\Pi}}_{k}=\hat{{p}}_{k}+e\hat{{\cal A}}(\hat{\vec{r}}_k)$ and (ii) a coupling to the electron spin through the Zeeman term $\hat{H}_k^{Z}=\frac{g^*\mu_BB}{2}\hat{\sigma}_{k,z}=\Delta \hat{\sigma}_{k,z}$ where $g^*$ is the effective electron gyromagnetic factor, $\mu_B$ is the Bohr magneton, and $\hat{\sigma}_{k,z}$ is the $z$-component Pauli matrix of electron $k$.

It is convenient to choose a new set of conjugate orbital coordinates, namely the center-of-mass (CM) coordinates given by $\vec{R}=\frac{1}{2}(\vec{r}_1+\vec{r}_2)$, $\vec{P}=\vec{p}_1+\vec{p}_2$ and the relative coordinates given by $\vec{r}=\vec{r}_1-\vec{r}_2$, $\vec{p}=\frac{1}{2}(\vec{p}_1-\vec{p}_2)$. We note that the e-e interaction only involves the relative position coordinate $\vec{r}$.

The system's Hamiltonian given in Eq.(\ref{Eq:q1}) transforms to the sum of a CM-Hamiltonian, a relative Hamiltonian and a Zeeman term:
\begin{eqnarray}
\hat{H}_{2e}=\hat{H}_{CM}+\hat{H}_r+\Delta \left ( \hat{\sigma}_{1,z}+\hat{\sigma}_{2,z} \right )
\label{Eq:qn2}
\end{eqnarray}
with the CM-Hamiltonian:
\begin{eqnarray}
\nonumber \hat{H}_{CM}=\frac{\hat{\vec{P}}^2}{2(2m^*)}+\frac{1}{2}(2m^*)\omega^2\hat{\vec{R}}^2+\frac{\omega_c}{2}\left ( \hat{R}_x \hat{P}_y- \hat{R}_y \hat{P}_x\right )\\
\label{Eq:qn3}
\end{eqnarray}
where $\omega_c=\frac{eB}{m^*}$ is the cyclotron frequency. The effective QD confinement is given by $\omega=\omega_0\sqrt{1+\left ( \frac{\omega_c}{2\omega_0} \right )^2}$.
The relative term in Eq.(\ref{Eq:qn2}) is given by:
\begin{eqnarray}
\nonumber \hat{H}_{r}&=&\frac{\hat{\vec{p}}^{\,2}}{2(\frac{m^*}{2})}+\frac{1}{2}(\frac{m^*}{2})\omega^2\hat{\vec{r}}^{\,2}\\
&+&\frac{\omega_c}{2}\left ( \hat{r}_x \hat{p}_y- \hat{r}_y \hat{p}_x\right )+
V_{e-e}(|\hat{\vec{r}}|)
\label{Eq:qn4}
\end{eqnarray}
For completeness, we start by defining the (dimensionless) CM annihilation operators
\[
\hat{\mathbf A}=\frac{1}{\sqrt{2}}\!\left(\frac{1}{\ell}\,\hat{\mathbf R}+\frac{i\,\ell}{\hbar}\,\hat{\mathbf P}\right),
\]
with corresponding creation operators $\hat{\mathbf A}^\dagger$, and, for the relative motion,
\[
\hat{\mathbf a}=\frac{1}{\sqrt{2}}\!\left(\frac{1}{2\ell}\,\hat{\mathbf r}+\frac{i\,2\ell}{\hbar}\,\hat{\mathbf p}\right),
\]
with creation operators $\hat{\mathbf a}^\dagger$. Here $\ell$ denotes the (magnetic-field–dependent) effective confinement length.
These operators satisfy canonical bosonic commutation relations; the full Cartesian derivation is given in Appendix~A.
We introduce here the bosonic chiral ladder operators employed throughout the text. For the center-of-mass motion we define
\[
A_R=\tfrac{1}{\sqrt{2}}(A_x-iA_y),\qquad A_L=\tfrac{1}{\sqrt{2}}(A_x+iA_y),
\]
and for the relative motion
\[
a_R=\tfrac{1}{\sqrt{2}}(a_x-i a_y),\qquad a_L=\tfrac{1}{\sqrt{2}}(a_x+i a_y).
\]
They satisfy canonical bosonic commutation relations, and their Hermitian conjugates act as creation operators; see Appendix~A for the full derivation in Cartesian coordinates. In terms of bosonic chiral operators (see Appendix A), the CM-Hamiltonian in Eq.(\ref{Eq:qn3}) can be rewritten as:
\begin{eqnarray}
\hat{H}_{CM}=\hbar \omega +\hbar \omega_R\hat{A}_{R}^{\dag}\hat{A}_{R}+\hbar \omega_L\hat{A}_{L}^{\dag}\hat{A}_{L}
\label{Eq:qn11}
\end{eqnarray}
with $B$-dependent chiral frequencies $\omega_R=\omega +\frac{\omega_c}{2}$ and $\omega_L=\omega -\frac{\omega_c}{2}$. Hence, the CM dynamics reduces to that of a two dimensional harmonic oscillator with different chiral right ($\hbar \omega_R$) and left ($\hbar \omega_L$) energy quanta. 
In a similar way, the relative Hamiltonian in Eq.(\ref{Eq:qn4}) transforms to:
\begin{eqnarray}
\hat{H}_{r}=\hbar \omega +\hbar \omega_R\hat{a}_{R}^{\dag}\hat{a}_{R}+\hbar \omega_L\hat{a}_{L}^{\dag}\hat{a}_{L}+V_{e-e}(|\hat{\vec{r}}|)
\label{Eq:qn12}
\end{eqnarray}
Thus, due to the presence of the e-e interaction term $V_{e-e}(|\hat{\vec{r}}|)$, the relative dynamics corresponds to that of a chirally {\it coupled} 2D oscillator. From now on, we will drop the trivial constant term $\hbar \omega$ in the expressions of both CM and relative Hamiltonians in Eqs.(\ref{Eq:qn11}-\ref{Eq:qn12}). 

The eigenstates and energies for the CM oscillators are given by $|N_R,N_L\rangle=\frac{1}{\sqrt{N_R!N_L!}}\hat{A}_{R}^{\dag\,N_R}\hat{A}_{L}^{\dag\,N_L}|0,0\rangle$ and $E_{N_R,N_L}^{(CM)}=\hbar\left ( N_R\omega_R+N_L\omega_L \right )$, with $N_R,N_L=0,1,2,...$. Additionally, the CM-angular momentum is given by $\hat{L}_z=\hbar\left ( \hat{A}_{R}^{\dag}\hat{A}_{R}-\hat{A}_{L}^{\dag}\hat{A}_{L}\right )$. Since $[\hat{H}_{CM},\hat{L}_z]=0$, the CM-angular momentum eigensystem $\hat{L}_z|N_R,N_L\rangle=\hbar\left ( N_R-N_L \right )|N_R,N_L\rangle$ is valid.

\subsection{\label{sec:level3} electron-electron interaction: \boldmath$\displaystyle \frac{1}{r^\eta}$}

The electron-electron interaction, which only affects the relative motion dynamics, is given in free space by the bare Coulomb repulsion potential $1/r$. Unlike free space, however, semiconductor QD systems are built from layers of different materials, each of which has its own dielectric properties -- which means that there will be image charges. As is well-known from textbook treatments of interacting dipoles etc., the presence of image charges creates an interaction which has a high power in $1/r$. Hence we consider in this paper a more realistic effective e-e interaction form $V_{e-e}(|{\vec{r}}|)\sim \frac{1}{r^{\eta}}$, with $1 \le \eta \le 2$. Coulomb repulsion corresponds to $\eta=1$, while $\eta=2$ represents a screened dipole-like interaction (Calogero model potential). 
Effective inverse-power interactions arise in several contexts (screening/image charges, collective-coordinate mappings, dimensionality reduction). Related derivations and applications for $1/r^s$ include the $1/r^2$ quantum breathing mode and Calogero dynamics~\cite{Date,Geller}, as well as exact treatments in semiconductor QDs~\cite{Namas}. See also recent nonequilibrium developments concerning Calogero-type models~\cite{Benjamin}.

The case of $\eta=2$ is not only justified physically due to  the presence of mirror charges in the semiconductor QD environment, it also allows for an analytical solution in the presence of an external magnetic field -- 
including mixing with all higher Landau levels as well as exactly including all exchange-correlation effects. By contrast, the $\eta=1$ (i.e. Coulomb repulsion) case only allows for exact solutions at a countable set of values for the confining potential \cite{Taut}. Furthermore, the resulting $\eta=2$ analytical results, energies and correlated states, can then be exploited for efficient numerical calculations for arbitrary $1 \le \eta \le 2$.

We start by focusing on the e-e potential  $V_{e-e}(|\hat{\vec{r}}|)= \frac{\tilde {\alpha}}{\hat{r}^2}$. As shown in Appendix A, $\hat{r}^2$ can be written in terms of chiral operators  as:

\begin{eqnarray}
\nonumber \hat{r}^2&=&2\ell^2\hat{\bar r}^2\\
&=&2\ell^2\left ( \hat{a}_{R}^{\dag}\hat{a}_{R}+\hat{a}_{L}^{\dag}\hat{a}_{L}+\hat{a}_{R}^{\dag}\hat{a}_{L}^{\dag}+\hat{a}_{R}\hat{a}_{L}+1\right )
\label{Eq:q13}
\end{eqnarray}
where a $B$-dependent effective confinement length is given by $\ell=\sqrt{\frac{\hbar}{2m^*\omega}}$.
Noting that $V_{e-e}(|\hat{\vec{r}}|)= \frac{\tilde{\alpha}}{\hat{r}^2}=2\hbar\omega\left (\frac{\alpha}{4\hat{\bar r}^2}\right )$, with a dimensionless e-e interaction strength $\alpha$, the relative Hamiltonian in Eq.(\ref{Eq:qn4}) can be written as:
\begin{eqnarray}
\hat{H}_{r}=\hbar \omega \left ( 2\hat{K}_z-1 \right )+\frac{\omega_c}{2} \hat{l}_z
\label{Eq:qn19}
\end{eqnarray}
where $\hat{K}_z=\frac{1}{2}\left ( \hat{a}_{R}^{\dag}\hat{a}_{R}+\hat{a}_{L}^{\dag}\hat{a}_{L}+1 \right )+\frac{\alpha}{4\hat{\bar r}^2}$ is one of the three generators of the dynamical $su(1,1)$ symmetry of the relative Hamiltonian, and the $r$-angular momentum is given by $\hat{l}_z=\hbar\left ( \hat{a}_{R}^{\dag}\hat{a}_{R}-\hat{a}_{L}^{\dag}\hat{a}_{L}\right )$ (see Appendices A-D).
Although $\hat K_z$ (via $1/r^2$) contains terms such as $\hat a_R\hat a_L$ and their Hermitian conjugates, the relative angular momentum operator $\hat l_z$ still commutes with the Hamiltonian $H_r$.
The explicit demonstration is given in Appendix~A1.

Since the relative angular momentum operator $\hat{l}_z$ turns out to be directly related to the Casimir operator, $\hat{l}_z$ does commute with every generator of the $su(1,1)$ Lie algebra, and consequently from Eq.(\ref{Eq:qn19}) it is evident that $[\hat{H}_{r},\hat{l}_z]=0$. Hence the eigenvalues $m\hbar$ of $\hat{l}_z$, with integer $m=-\infty,...,\infty$, are good quantum numbers. The relative Hamiltonian eigenvalue equation is now expressed as $\hat{H}_{r} |n,m\rangle_{\alpha}=E_{n}(\alpha,m)|n,m\rangle_{\alpha}$ with $n=n_{min},...,\infty$, where $n_{min}=m\Theta (m)$ ($\Theta (x)$ the Heaviside function) and $\hat{l}_z|n,m\rangle_{\alpha}=\hbar m|n,m\rangle_{\alpha}$ (integer $m=-\infty,...,\infty$) \cite{qaj}. Under the action of an applied magnetic field, the ground state corresponds to $m \le 0$, hence $n_{min}=0$.

From the $su(1,1)$ dynamical structure, it follows that the relative motion Hamiltonian has a discrete energy spectrum given by:
\begin{eqnarray}
E_{n}(\alpha,m)&=&E_{0}(\alpha,m)
+n\,2\hbar \omega
\label{Eq:qq30}
\end{eqnarray}
where $E_{0}(\alpha,m)$ denotes the lowest energy level associated with a given $m$-subspace. Notice that for a given $m$, the only effect of the e-e interaction is to fix the lowest energy level whereas consecutive energy levels are separated by $2\hbar \omega$, regardless of the e-e interaction strength $\alpha$. Notice that the CM motion, as governed by Eq.(\ref{Eq:qn11}), is identical to the non-interacting relative motion, i.e. $\alpha=0$.

For every $m$-sector, the lowest energy state $|0,m\rangle_{\alpha}$ must satisfy:
\begin{eqnarray}
\hat{K}^-|0,m\rangle_{\alpha}=0
\label{Eq:q28}
\end{eqnarray}
or equivalently
\begin{eqnarray}
\hat{\bar r}^2\hat{k}^-|0,m\rangle_{\alpha}=\frac{\alpha}{4} |0,m\rangle_{\alpha}
\label{Eq:q29}
\end{eqnarray}
which means that $|0,m\rangle_{\alpha}$ is an eigenstate of the operator $\hat{\bar r}^2\hat{k}^-$ with eigenvalue $\frac{\alpha}{4}$.
We now consider the representation of the interacting states $|n,m\rangle_{\alpha}$ in terms of the non-interacting states $|n_R,n_L\rangle$.

As fully detailed in Appendix B, the interacting $m$-ground state $|0,m\rangle_{\alpha}$, expressed in the basis of chiral oscillator eigenstates $|n_R,n_L\rangle$, takes the following form:
\begin{eqnarray}
\nonumber |0,m\rangle_{\alpha}=\sqrt{|m|!}\frac{\Gamma\left ( \frac{\alpha_m+m}{2}+1 \right )}{\sqrt{\Gamma \left ( \alpha_m+1 \right )}}L_{\frac{\alpha_m+m}{2}}^{-m}\left ( -\hat{k}^{+}\right )|0,|m|\rangle\\
\label{Eq:q36}
\end{eqnarray}
where $L_{\nu}^{-m}(-\hat{k}^{+})$ denotes a Laguerre function with the operator $\hat{k}^{+} = \hat{a}_{R}^{\dag} \hat{a}_{L}^{\dag}$ as its argument, $\Gamma(z)$ represents the Gamma function, and $\alpha_m = \sqrt{m^2 + \alpha}$. The corresponding interacting ground state energy $E_{0}(\alpha,m)$ is given by:

\begin{eqnarray}
E_{0}(\alpha,m)=\hbar \omega \alpha_m+m \frac{\hbar \omega_c}{2}
\label{Eq:q37}
\end{eqnarray}
in full agreement with previously reported results \cite{qaj}. 
Inserting this last result in Eq.(\ref{Eq:qq30}), the QD full energy spectrum can be identified. 

The complete set of interacting relative states is given by applying powers of the raising generator  $\hat{K}^{+}$ operator to the lowest state of every $m$-sector:
\begin{eqnarray}
|n,m\rangle_{\alpha}&=&\sqrt{\frac{\Gamma(\alpha_m+1)}{\Gamma(\alpha_m+n+1)}}\, \frac{\hat{K}^{+ n}}{\sqrt{n!}}|0,m\rangle_{\alpha}
\label{Eq:b133}
\end{eqnarray}
\\

For an effective e-e repulsion term with $\eta < 2$, we use the exact analytical results previously obtained for $\eta=2$ as the basis for searching numerically solutions for the general $\eta$ case. Hence, we write the relative Hamiltonian in Eq.(\ref{Eq:qn19}) as follows:

\begin{eqnarray}
{\hat H}_r= \hbar \omega \left ( 2\hat{K}_z-1 \right )+\frac{\omega_c}{2} \hat{l}_z+\hbar \omega\frac{\alpha}{2}\left ( \frac{2^{1-\frac{\eta}{2}}}{\hat{\bar r}^{\eta}}-\frac{1}{\hat{\bar r}^{2}} \right )
\label{Eq:tl77}
\end{eqnarray}
where $V_{e-e}(|\hat{\vec{r}}|)=\hbar \omega\frac{2^{-\frac{\eta}{2}}\alpha}{\hat{\bar r}^{\eta}}$ for general $\eta$ and $\hat{\bar r}$ is given by Eq.(\ref{Eq:q13}).

\section{\label{sec:level3}Interaction with Twisted Light}
Light with orbital angular momentum, also known as twisted light, can transfer its angular momentum to trapped electrons or fermion atoms in a controlled way.
TL with circular polarization (${\hat e}_{\phi}=\frac{1}{\sqrt{2}}\left (  {\hat e}_{x}+i{\hat e}_{y} \right )$), propagating in the $z$-direction, can be described by the vector potential \cite{quinteiro2022,jifuksprb2023}:
\begin{eqnarray}
{\vec A}({\vec r},t)=e^{-i\left ( \omega_T t-kz \right )}\left [ F_{k,l}\left ( r \right )e^{i l \phi} \right ]{\hat e}_{\phi}+c.c.
\label{Eq:tl1}
\end{eqnarray}
where $ F_{k_r,l}=A_0J_{l}(k_r r)$ describes the radial profile of the TL beam ($J_{l}(\rho)$ is a Bessel function) with $1/k_r$ the beam waist and $\omega_T$ the frequency of the light. In the cases we will be considering below, $k_r r<<1$ so that the approximation $F_{k_r,l}\approx \frac{(k_r r)^l}{2^ll!}$ holds. The parameter $l$ denotes the vorticity of the TL, which reflects
the intrinsic topological characteristics of the beam and plays a crucial role in understanding
the structure and properties of the vortex light field.
In paraxial optics, Laguerre--Gauss (LG) and Hermite--Gauss (HG) beams are the natural transverse eigenmodes of stable laser resonators and are routinely produced experimentally, whereas ideal nondiffracting Bessel beams provide an analytically convenient profile. 
We adopt the Bessel form to keep closed-form expressions and to make explicit the azimuthal phase $e^{i l\phi}$ that governs OAM transfer and the ensuing selection rules. 
In the near-axis/small-cone regime (Bessel: $k_r r\ll1$; LG: $r\ll v_0$, where $v_0$ denotes the LG  spot size, in length units), the $l=1$ components satisfy
$J_1(k_r r)\simeq (k_r r)/2$ and $\mathrm{LG}_{p=0}^{\,l=1}(r)\propto (r/v_0)\,e^{-r^2/v_0^2}\simeq (r/v_0)\,[1-\mathcal O(r^2/v_0^2)]$, 
so both fields reduce to $r\,e^{i\phi}$ multiplied by a slowly varying radial envelope. 
Therefore, the quadrupolar coupling and the selection rule $\Delta m=\pm2$ apply equally to LG beams; HG beams can be decomposed into LG modes, so any $l=1$ component yields the same selection rule. 
Different radial envelopes (Gaussian for LG, Cartesian nodes for HG) only rescale matrix elements via radial overlaps and hence affect intensities, not the allowed character of the transitions.

A single electron couples with a TL pulse, in the length gauge, by the interaction Hamiltonian 
$\hat{H}_{TL} = -\frac{q_e}{l+1}\, \vec{r}_\perp \cdot \frac{\partial\, \vec{A}(\vec{r}, t)}{\partial t},
$
where \( q_e \) is the electron charge.

Thus, the interaction of a two-electron system with a TL pulse should be described by a Hamiltonian as follows:
\begin{eqnarray}
\nonumber \hat{H}^{(l)}_{2e-TL}(t) &=& -i\left( \frac{\omega_T q_e A_0}{l+1} \right)\frac{k_r^l}{2^l l!}
f(t) \\
&\times& \left[ e^{-i\omega_T t} \sum_{j=1,2} \hat{\rho}_j^{l+1} + e^{i\omega_T t} \sum_{j=1,2} \hat{\rho}_j^{\dag\,l+1} \right],
\label{Eq:tl-mod}
\end{eqnarray}
where \( f(t) = \frac{1}{\sqrt{2\pi \tau^2}} e^{-t^2 / (2\tau^2)} \) is a normalized Gaussian envelope describing a finite pulse of length $\tau$ and \( \hat{\rho}_j = \hat{x}_j + i \hat{y}_j \) is the complex in-plane position operator of the \( j \)-th electron.

Throughout this work, we consider excitation in the near-resonant regime \( \omega_T \sim \omega_0 \), with sufficiently long pulses such that \( \omega_T \tau \approx 1 \).

For $l=0$, the TL reduces to a non-vortex light and the usual electric dipole coupling is retrieved:

\begin{eqnarray}
\nonumber {\hat H}^{(l=0)}_{2e-TL}&=&i\left (2q_eA_0\omega_T\right )f(t)\left [ e^{-i\omega_T t}\left ( \hat {R}_x+i \hat {R}_y\right )-H.C.  \right ]\\
\nonumber &=&i\left (2q_e\ell_0 A_0\omega_T\right )f(t)\left [ e^{-i\omega_T t}\left ( \hat{A}_{L}+\hat{A}_{R}^{\dag}\right )-H.C. \right ]\\
\label{Eq:tl3}
\end{eqnarray}

This last result emphasizes the Kohn's theorem, i.e. an unstructured light pulse only couples with the CM degree of freedom remaining dark the correlation e-e effects. In Eq.(\ref{Eq:tl3}) $H.C.$ denotes hermitian conjugate.

However for $l=1$, an electric quadrupole coupling of TL with both CM and relative degrees of freedom takes place:

\begin{eqnarray}
\nonumber {\hat H}^{(1)}_{2e-TL}&=&
i\left ( \frac{q_eA_0k_r\ell_0^2 \omega_T}{2}\right )f(t)\\
\nonumber &\times&\left \{ e^{-i\omega_T t}\left [ \left (\hat{A}_{R}^{\dag}+\hat{A}_{L} \right )^2+ \left (\hat{a}_{R}^{\dag}+\hat{a}_{L} \right )^2 \right ]-H.C.\right \}\\
\label{Eq:tl4}
\end{eqnarray}

We note that TL with $l=1$ affects identically CM and relative motion separately -- however it does not mix them. Incidentally, the presence of the TL enlarges the system's dynamical Lie algebra from the previous $su(1,1)$ matter Lie algebra to a larger $10$-generator $so(3,2)$ Lie algebra.

\section{\label{sec:level3a}Results}

To illustrate the behavior of the system, we focus on a prototypical GaAs quantum dot of two interacting electrons. The effective mass and g-factor are taken as \( m^* = 0.067\,m_0 \) and \( g^* = -0.44 \), respectively. Energies are given in units of \( \hbar \omega_0 \). 
The relative-motion sector for the Coulomb case is expanded in the exact Calogero eigenbasis $\{\lvert n,m\rangle\}$; we typically employ up to $n_{\max}=30$ radial excitations per $m$-sector and include $m$ values sufficient to cover the ground-state and the $\Delta m=\pm2$ optical channels over the magnetic-field range considered (practically, $\lvert m\rvert\le 30$). 
This truncation yields converged spectra and absorption; increasing to $n_{\max}=40$ changes excitation energies and integrated absorption by less than $10^{-3}$. The TL-driven response is iterated in time until successive updates of the relevant observables fall below $10^{-6}$, at which point we take this as the stationary solution.

The above results establish a rigorous platform from which the impact of particle interactions can be explored, specifically Eq.(\ref{Eq:tl77}).

\subsection{\label{sec:level4}{Twisted light's chiral quanta excitations}}
From now on, we focus on the TL pulse absorption by the system's relative motion, since it represents a new scenario beyond the range of validity of Kohn's theorem. The resulting Hamiltonian is:
\begin{eqnarray}
\nonumber
{\hat H}^{(l=1)}_{2e}&=& \hbar \omega_R\hat{a}_{R}^{\dag}\hat{a}_{R}+\hbar \omega_L\hat{a}_{L}^{\dag}\hat{a}_{L}+V_{e-e}(|\hat{\vec{r}}|)\\
\nonumber&+&
i{\cal E}_{0}f(t)\left [ e^{-i\omega_T t} \left (\hat{a}_{R}^{\dag}+\hat{a}_{L} \right )^2 -e^{i\omega_T t} \left (\hat{a}_{R}+\hat{a}_{L}^{\dag} \right )^2   \right ]\\
\label{Eq:tl7}
\end{eqnarray}
where the dimensionless light-matter coupling strength is ${\cal E}_{0}= \frac{q_eA_0k_r\ell_0^2\omega_T}{2}$. According to Eq.(\ref{Eq:tl7}) the quadrupole-like TL-matter coupling satisfies the selection rule $\Delta m=\pm 2$.

Therefore, the matter state remains in the same parity sector and consequently the two-electron spin state is unchanged by TL absorption: hence spin effects are irrelevant for the dynamics in the present case.

For non-interacting electrons, i.e. $V_{e-e}(|\hat{\vec{r}}|)=0$, a two quanta excitation resonance occurs at $\omega_T = 2\omega_R$. Taking a rotating-wave-approximation in the interaction picture yields the simplified Hamiltonian (a similar expression holds for CM):
\begin{eqnarray}
{\hat H}^{(1)}_{2e-TL}=i{\cal E}_{0}f(t)\left ( \hat{a}_{R}^{\dag\, 2}-\hat{a}_{R}^{2} \right )
\label{Eq:tl6}
\end{eqnarray}
which is a well-known squeezing Hamiltonian in quantum optics.

\begin{figure}[ht]
    \centering
   \includegraphics[width=0.45\textwidth]{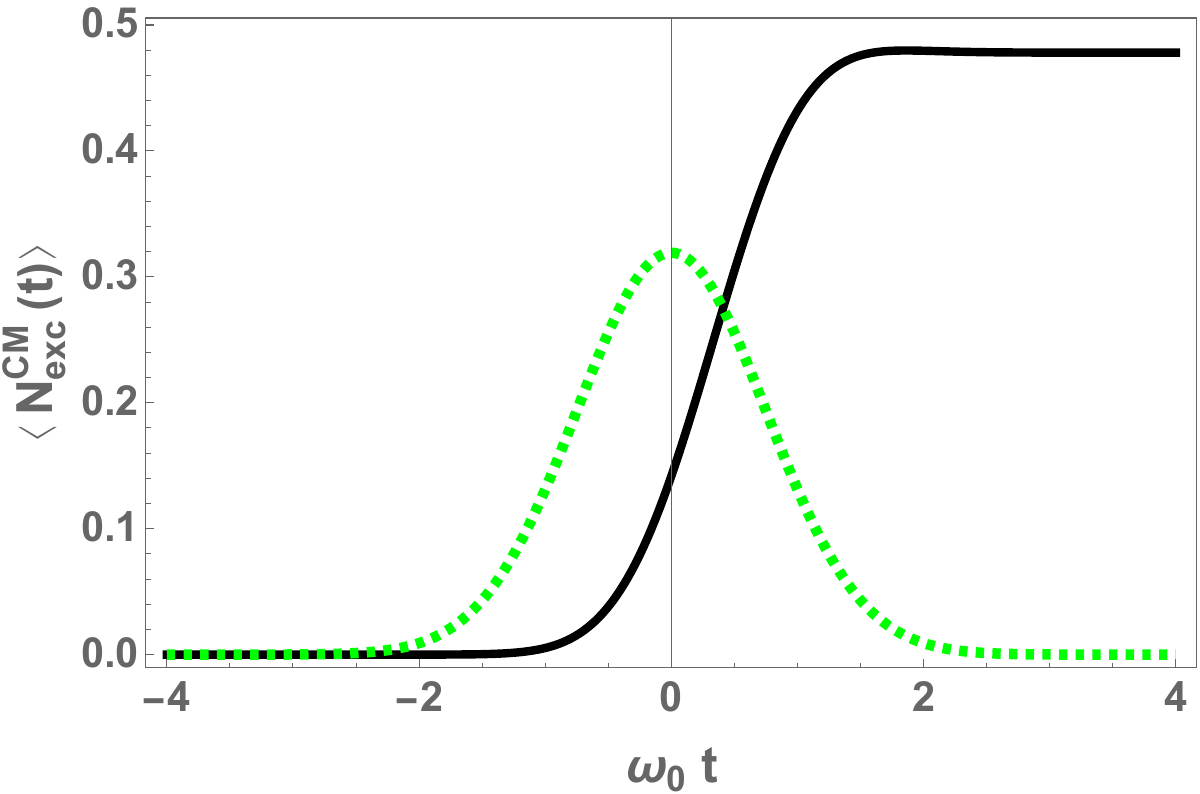}
    \caption{Time evolution of the CM mean number of chiral excitations (solid black line) in response to a Gaussian pulse for $\omega_0\tau=$1 and amplitude pulse $0.1$ (green broken line, arbitrary units).}
    \label{fig:pulseexcitation}
\end{figure}

The mean number of right chiral relative excitations, $\langle \hat{a}_{R}^{\dag}\hat{a}_{R}\rangle$, can be calculated analytically yielding:
\begin{eqnarray}
\langle \hat{a}_{R}^{\dag}\hat{a}_{R}\rangle(t)={\rm Sinh}^{2}\left [ \frac{{\cal E}_0}{2}\left ( 1+ Erf \left ( \frac{t}{\sqrt{2}\tau} \right ) \right ) \right ]
\label{Eq:tl97}
\end{eqnarray}

while $\langle \hat{a}_{L}^{\dag}\hat{a}_{L}\rangle=0$. $Erf(x)$ denotes the error function.  Fig.~\ref{fig:pulseexcitation} shows this result for a Gaussian pulse. Notice that this non-interacting case produces identical results to the center-of-mass (CM) case.

Hence, we write the relative-TL Hamiltonian as:
\begin{eqnarray}
{\hat H}^{(1)}_{2e}= {\hat H}_r
+
i{\cal E}_{0}f(t)\hat{\bar r}^{2}\left [ e^{-i\omega_T t} e^{2i \phi}  -e^{i\omega_T t} e^{-2i \phi}   \right ]
\label{Eq:tl177}
\end{eqnarray}
with ${\hat H}_r$ given by Eq.\eqref{Eq:tl77}), where we have made use  of $\left ( {\hat r}_x\pm i {\hat r}_y \right )^2=\ell_0^2\hat{\bar r}^{2}e^{\pm 2i \phi}$ which reinforces the fact that TL with vorticity number $l=1$ induces transitions involving two chiral quanta transitions, i.e the QD ground state is mainly coupled to the lowest energy state corresponding to an angular momentum variation $\Delta m=-2$ (a real space representation can be useful as detailed in Appendix E).
To obtain the time evolution of the
mean excitation number of relative chiral quanta for the QD-TL system, we solve numerically the full time-dependent Schr\"odinger equation with driving Hamiltonian \eqref{Eq:tl77} in the basis $\{ |n,m\rangle_{\alpha} \}$ (see Eq.\eqref{Eq:b133}). Since the TL absorption should be proportional to the asymptotic long- time limit of the number of QD chiral excitations generated by the TL pulse, i.e.  $\langle N_{exc} \rangle \left ( t\rightarrow \infty\right )$, from now on we focus on that quantity.

\subsection{\label{sec:level7}{Twisted light absorption: magnetic field effects}}

We define the mean number of chiral excitations, given by \( \langle N_{\text{exc}} \rangle = \langle \hat{a}_{R}^{\dag}\hat{a}_{R} + \hat{a}_{L}^{\dag}\hat{a}_{L} \rangle \), as a quantitative measure of twisted light absorption by the two-electron quantum dot system. The net number of chiral excitations generated by the TL pulse is defined as \( \Delta N_{\text{exc}} = \langle N_{\text{exc}}(\infty) \rangle - \langle N_{\text{exc}}(-\infty) \rangle \), where \( \langle N_{\text{exc}}(-\infty) \rangle \) accounts for the mean number of chiral excitations present in the interacting ground state prior to the arrival of the TL pulse. In the non-interacting case, corresponding to \( \alpha = 0 \), the ground state contains no chiral excitations and hence \( \langle N_{\text{exc}}(-\infty) \rangle = 0 \).

The TL driving term \( (\hat{a}_R^\dagger + \hat{a}_L)^2 \) (and corresponding H.C. term) in Eq.(\ref{Eq:tl7}), preserves spin symmetry and thus selectively couples states with the same spin parity. As a result, singlet spin states,
can only couple among themselves, while spin triplet states, i.e. odd \( m \) values, 
form a separate excitation ladder. Thus, the absorption of TL with vorticity $l=1$ is connected directly to the generation/annhilation of pairs of right/left chiral excitations.

The TL induced transition probabilities between the relative motion eigenstates, depend on the TL detuning with respect to the lowest electronic energy levels.

\begin{figure}[htbp]
    \centering
    \begin{subfigure}[b]{0.51\textwidth}
        \includegraphics[width=\textwidth]{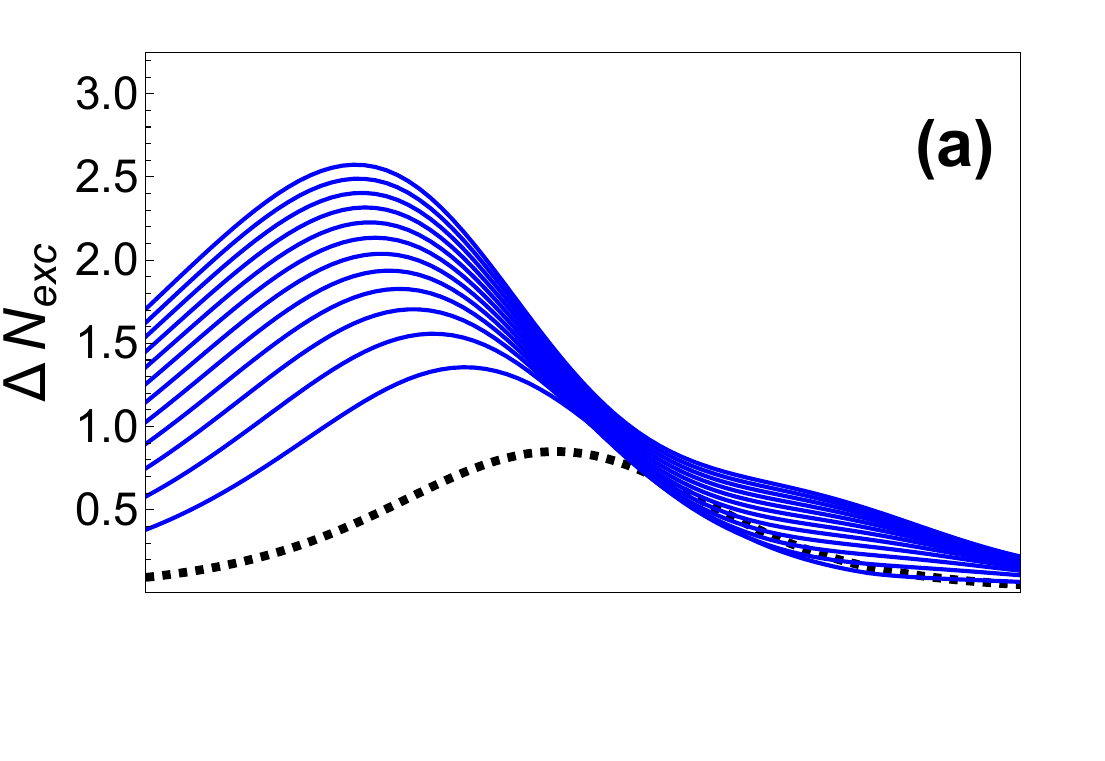}
        \label{fig:sub-coul}
    \end{subfigure}
    \begin{subfigure}[b]{0.51\textwidth}
    \vspace{-2.24cm}
        \includegraphics[width=\textwidth]{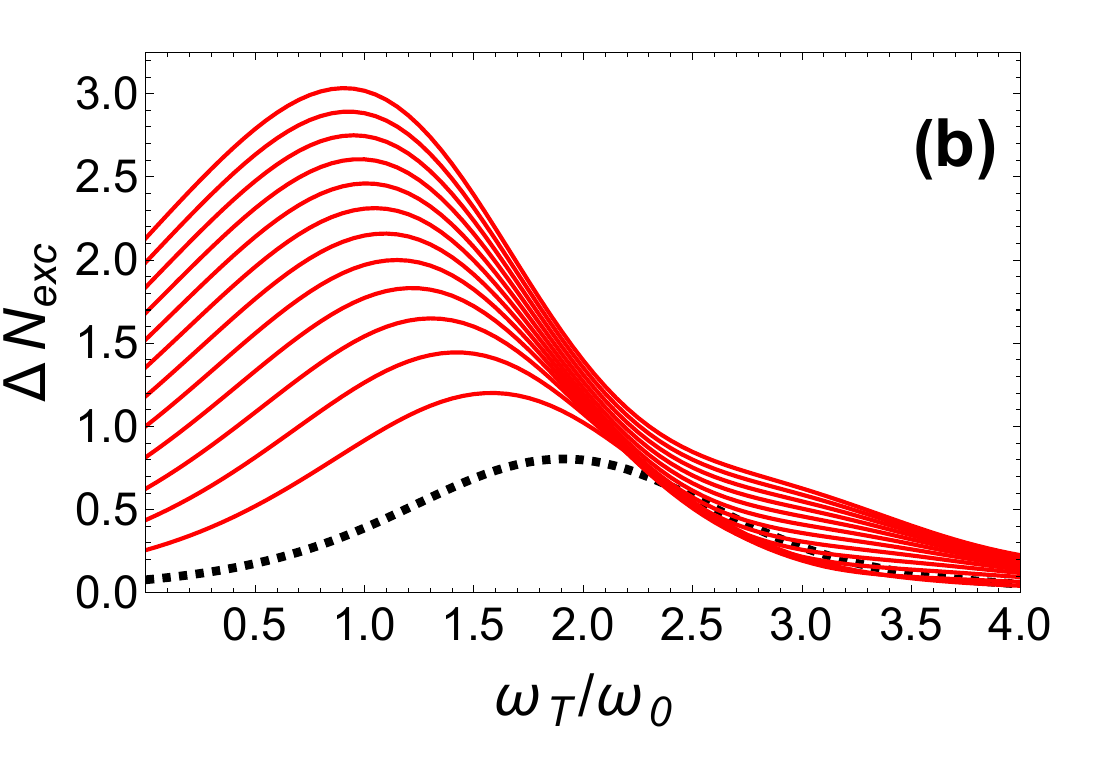}
        \label{fig:sub-calo}
    \end{subfigure}
    \caption{Net mean number of relative chiral excitations $\Delta N_{\text{exc}}$ as a function of the TL pulse central frequency. Black dashed curve: non-interacting case or equivalently CM case. As the interaction strength increases, a double absorption peak develops. The curves increase from bottom to top in steps of 0.5 for $0.0 \leq \alpha \leq 6$. (a) Calogero (b) Coulomb.}
    \label{fig:calo-coul}
\end{figure}

The above results establish a rigorous platform from which the interplay of particle interactions and vortex light can be explored – specifically
Eq.(\ref{Eq:tl7}) or equivalently Eq.(\ref{Eq:tl177}). Though exploring this fully for different types of functional interaction forms remains a job for future work, the formal
framework that we have already developed in the previous section allows us to assess what will happen.
We start our analysis by considering the zero magnetic field case, i.e. $\omega_c=0$. Fig. \ref{fig:calo-coul} shows $\Delta N_{\text{exc}}$, the asymptotic mean value of chiral excitations as a rough measure of TL absorption, as a function of the pulse TL central frequency
$\omega_T/\omega_0$. For non-interacting electrons, i.e. $\alpha = 0$, a single absorption peak centered at $\omega_T/\omega_0=2$ is observed (black dashed line),
which is also identical to the case for the CM response. 

As the e-e interaction increases, this peak moves to lower values
of the pulse central frequency and simultaneously becomes much higher in magnitude than the CM case, hence indicating a
strongly enhanced TL absorption induced by the e-e interaction. This finding contradicts the prediction of Kohn’s theorem. The emergence and behavior of this peak can be understood as the net result of several effects: (1) the
energy splitting between the ground state at $m = 0$ and the lower energy state at $m = -2$ which decreases as $\alpha$ goes larger;
and (2) the TL associated transition matrix element between the ground state at $m = 0$ and the second energy state at $m = -2$
which, although vanishing for the non-interacting case, starts to grow as $\alpha$ increases, while at the same time the energy splitting
decreases as a result of the competition between the harmonic confinement, which squeezes the particles near the origin, and
the e-e interaction which pushes the electrons further away from one another.

\begin{figure}[htbp]
    \centering
        \includegraphics[width=0.5\textwidth]{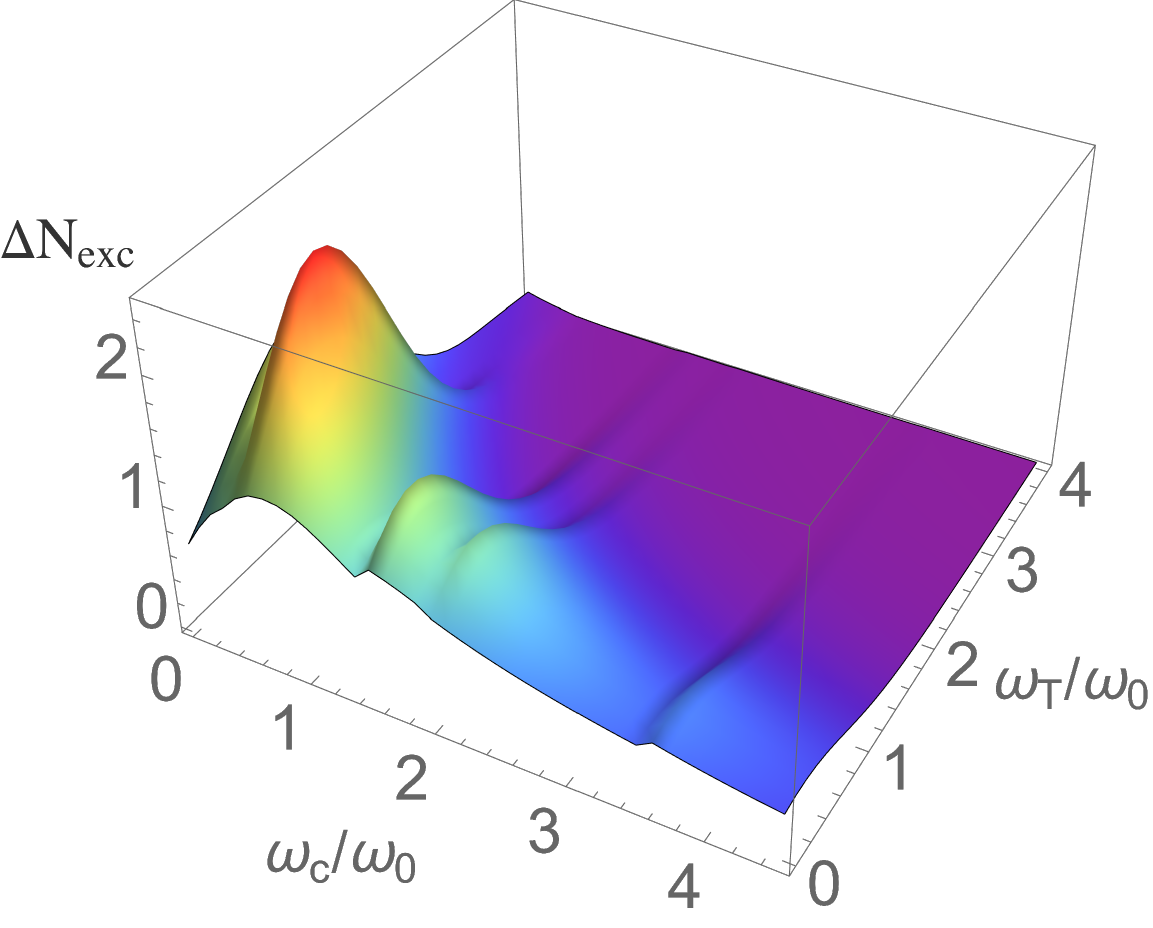}
\caption{ The net change in the number of chiral excitations, \( \Delta N_{\text{exc}} = \langle N_{\text{exc}}(\infty) \rangle - \langle N_{\text{exc}}(-\infty) \rangle \), as a function of the TL pulse central frequency. At zero magnetic field and for non-interacting electrons, the dominant absorption peak is centered near \( \omega_T/\omega_0 = 2.0 \), identical to the CM case. As the magnetic field increases, a rich interplay with the interaction strength emerges, indicating ground-state parity switching detectable via TL absorption. The results correspond to the Calogero interaction model with coupling parameter \( \alpha = 1.25 \).
}
     \label{fig:nexcwlwc}
\end{figure}

We proceed by considering finite magnetic field effects. From now on we fix $\alpha=1.25$ for the e-e dimensionless interaction strength. In figure~\ref{fig:nexcwlwc} we show the chiral excitation mean number \( \Delta N_{\text{exc}} \) as a function of both the magnetic field \( \omega_c/\omega_0 \) and the TL central frequency \( \omega_T/\omega_0 \). This observable reflects the net angular momentum transferred from the twisted light to the two-electron quantum dot. The response landscape exhibits a series of resonant peaks whose structure can be understood by examining the energy spectrum as a function of the relative angular momentum \( m \), as shown in the inset of Fig.~\ref{fig:fig2} in Appendix F.

At low magnetic fields (e.g., \( \omega_c/\omega_0 = 0.10 \)), the QD ground state lies in the singlet sector with \( m = 0 \), and the nearest accessible spin singlet state with \( m = -2 \) is still near $\omega_T/\omega_0=2$.

As a result, absorption remains weak across the TL low frequency range. As the field increases, the energy ladder for $m<0$ tilts downward, progressively favoring transitions at lower frequencies toward the singlet sector with \( \Delta m = -2 \), giving rise to a pronounced excitation peak in Fig.~\ref{fig:nexcwlwc} around \( \omega_c/\omega_0 \approx 0.15 \).

Nevertheless, for \( \omega_c/\omega_0 \approx 0.38 \), the ground state switches again to a triplet spin state. Around \( \omega_c/\omega_0 \approx 1.00 \), the energy separation between the low-lying triplet states is significantly reduced, making transitions within the triplet sector energetically favorable at still lower TL frequencies. This mechanism accounts for the sharp excitation peaks observed at \( \omega_T/\omega_0 \approx 0.45 \).

At higher fields (e.g., \( 1.40 < \omega_c/\omega_0 < 1.90 \)), the QD ground state goes back to the singlet sector with \( m = -2 \). In this narrow magnetic field window, the resonance condition is again satisfied, and a moderate chiral excitation signal is recovered. Secondary features near \( \omega_T/\omega_0 \approx 2 \) correspond to off-resonant transitions to higher-energy singlet states allowed only for interacting electrons, i.e. $\alpha \ne 0$.

Furthermore, at high magnetic fields the QD ground state freezes in the triplet spin sector, the energy mismatch between the available levels exceeds the spectral width of the optical pulse, making transitions increasingly off-resonant. As a result, the system fails to absorb the incoming light and \( \Delta N_{\text{exc}}(t) \) is vanishing small. 

\begin{figure}[htb]
\centering
\includegraphics[width=0.45\textwidth]{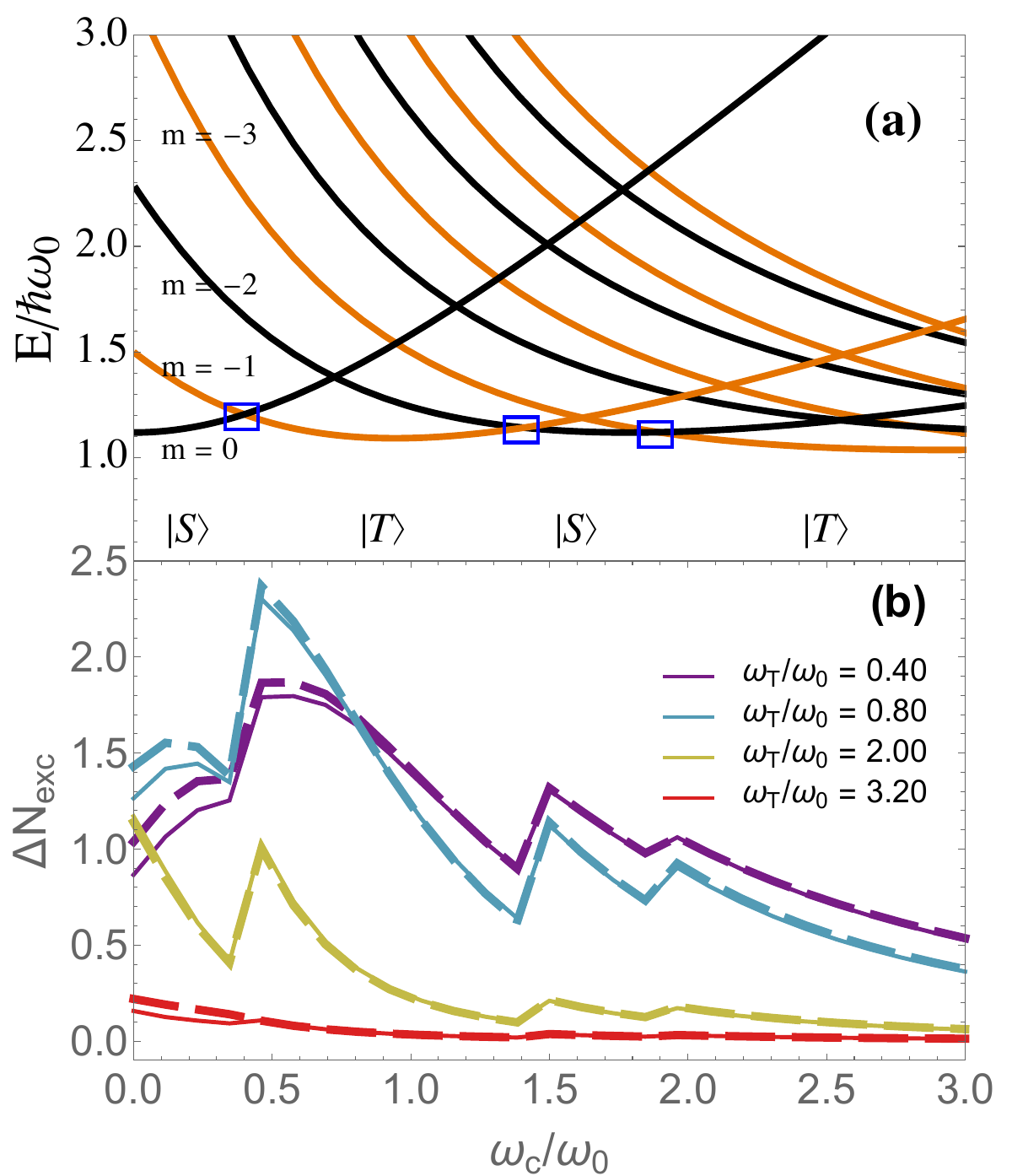}

\caption{(a) Energy levels of a two-electron quantum dot under an external magnetic field. Parity transitions between
$|S>$ and $|T>$ spin states are indicated by square markers.
(b) Chiral excitation number as a function of light frequency, illustrating ground-state parity switching. Solid lines correspond to the Coulomb interaction, while dashed lines represent the Calogero model with interaction strength 
$\alpha=$ 1.25.} 
    \label{fig:combined}
\end{figure}

In order to explore finely how e–e interactions influence the TL absorption in two-electron quantum dots Figures~\ref{fig:combined}(a) and (b) depict the many-body energy spectrum and the corresponding excitation response as a function of the normalized magnetic field \( \omega_c/\omega_0 \), for several fixed values of the TL frequency \( \omega_T/\omega_0 \). In panel (a), eigenstates are labeled according to their spin sector—singlet (\( |S\rangle \)) or triplet (\( |T\rangle \))—and blue squares highlight magnetic field values at which the ground state undergoes spin symmetry changes. 

These transitions are clearly reflected in Fig.~\ref{fig:combined}(b), where the excitation spectrum exhibits distinct step-like features in \( \Delta N_{\text{exc}} \). Each discontinuity signals a sudden change in the spin symmetry, driven by the e-e interaction on the ground-state spatial configuration, that enhance or suppress the interaction with twisted light. 
Obviously, this jump behavior is absent in the non-interacting case, where the excitation profile remains smooth and monotonic (see Fig.~\ref{fig:calo-coul2}).
Moreover, the use of fixed-frequency light in combination with a magnetic-field sweep provides \cite {Deric} a practical and sensitive method for detecting spin-sector and parity transitions. Twisted light selectively couples to \( \Delta m = \pm 2 \) channels, and the resulting chiral excitation signal acts as a clear spectroscopic marker of symmetry changes—even those occurring within narrow field intervals.

A key observation is that the TL absorption resulating from the excitation spectra computed using Coulomb and Calogero interactions are nearly indistinguishable, confirming that the Calogero model accurately reproduces the magneto-optical response of the system. The main effect of using the Calogero interaction is a global shift in the energy levels, as shown in the Appendix F (Fig.~\ref{fig:fig2}), while preserving the structure of level crossings and symmetry transitions. This makes the Calogero model a highly efficient and reliable framework for capturing the essential physics of two-electron systems in magnetic fields.

\subsection{\label{sec:level7a}{Center-of-mass and relative magneto-optical response to TL}}
\begin{figure}[htbp]
    \centering
    \begin{subfigure}[b]{0.50\textwidth}
       \includegraphics[width=\textwidth]{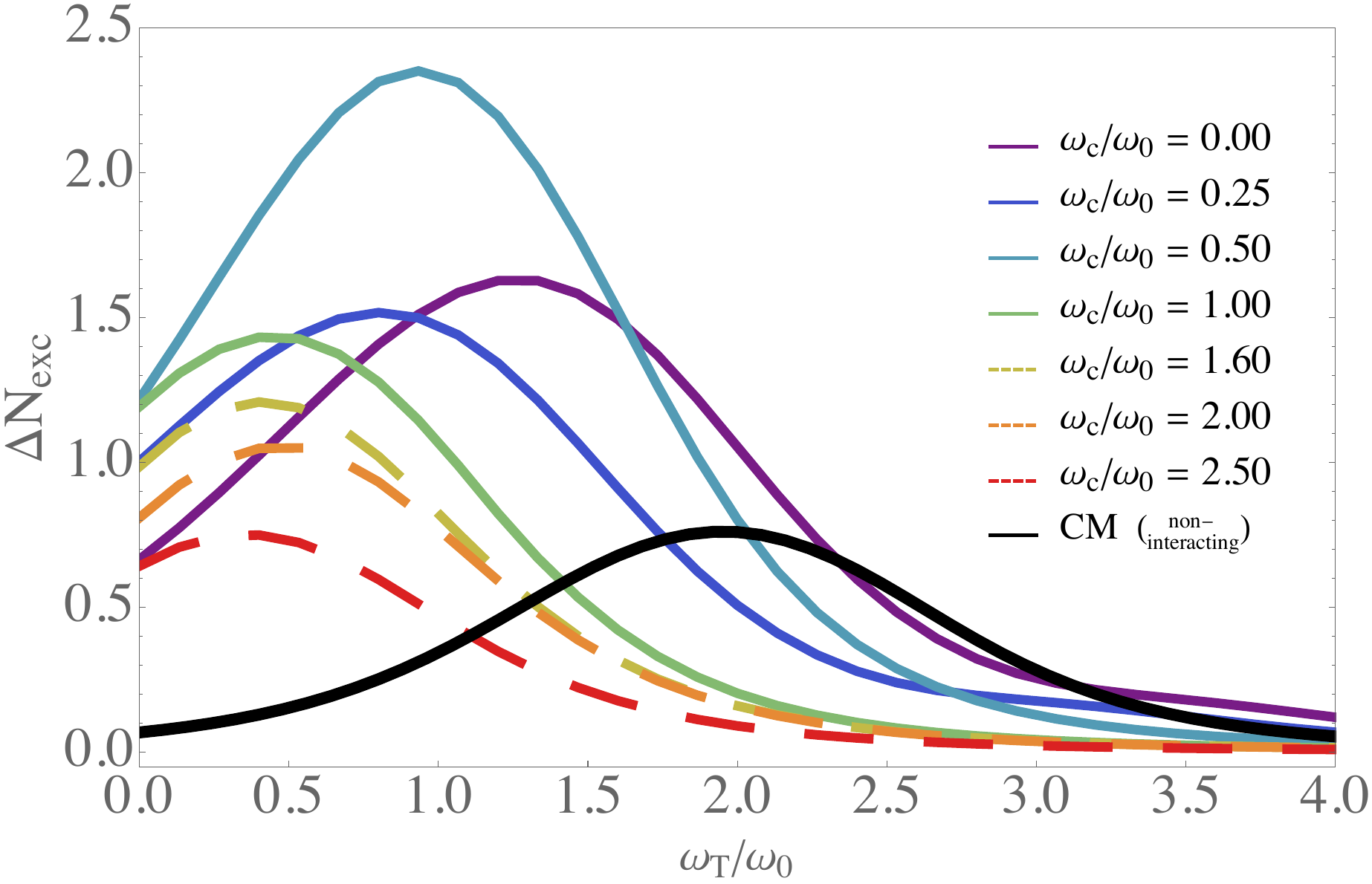} 
    \end{subfigure}
\caption{The net change in the number of chiral excitations, \( \Delta N_{\text{exc}} = \langle N_{\text{exc}}(\infty) \rangle - \langle N_{\text{exc}}(-\infty) \rangle \) as a function of the TL pulse central frequency. At zero magnetic field and for non-interacting electrons, the dominant absorption peak is centered at $\omega_T/\omega_0 \approx 2.0$, identically to the CM case. As the magnetic field strength increases, a rich interplay with the e-e interaction strength emerges, indicating ground-state parity switching detected by TL absorption.}
\label{fig:calo-coul2}
\end{figure}

For non-interacting electrons, the 2-electron-QD excitations can be constructed from the one-electron levels, provided that the spin–orbital symmetry of the lowest pair is preserved. These states are non-degenerate due to Zeeman splitting, since their energetic configuration is shaped by the magnetic field. Accordingly, in the non-interacting regime, the total number of excitations matches that of the center-of-mass motion.
Figure~\ref{fig:calo-coul2} illustrates a remarkable transformation in the excitation response of the system when electron-electron interactions and magnetic field are taken into account. In the non-interacting e-e case, equivalent to the center-of-mass case, as depicted by the black curve in Fig.~\ref{fig:calo-coul2}, the excitation profile exhibits a broad resonance centered around \( \omega_T/\omega_0 \approx 2 \), consistent with the Kohn theorem, which guarantees that the CM motion remains unaffected by interactions under harmonic confinement. This resonance arises because the CM behaves as a two-dimensional harmonic oscillator, and the optical field used in this case is not a conventional dipolar excitation (which would couple linearly to the position operators \( \hat{x}, \hat{y} \)). Instead, it is a \textit{structured light field carrying orbital angular momentum} (vorticity), which couples \textit{quadratically} to the position operators, specifically, to combinations such as \( \hat{x}^2 - \hat{y}^2 \) and \( 2\hat{x}\hat{y} \). These correspond to operators of the form \( (\hat{A}_R^\dagger + \hat{A}_L)^2 \), which induce transitions with \( \Delta m = \pm 2 \)  and leads to a resonance condition when the driving frequency matches \textit{twice} the oscillator frequency, i.e., \( \omega_T = 2 \omega_0 \). Beyond this excitation, the CM remains essentially transparent to the optical field, as no further channels are efficiently accessible within the dipole-quadratic approximation.

In stark contrast to CM excitations, the relative-motion spectrum is highly sensitive to the e-e interaction. As a result, the chiral excitation response becomes significantly enhanced, both in intensity and spectral structure. Not only does the number of excitations \( \Delta N_{\text{exc}} \) increase markedly compared to the CM case, but the resonance peaks also undergo a systematic redshift, shifting toward lower values of \( \omega_T \). This behavior originates from magnetic-field-induced parity changes in the ground state, as illustrated in Fig.~\ref{fig:combined}(a) and inset in Fig.~\ref{fig:fig2}.

While the Zeeman term shifts the position of spin-sector crossings by modifying the ground-state energy hierarchy, it does not alter the structure of the excitation spectrum. The spectral shape and selection rules remain governed entirely by the twisted-light coupling, which determines the allowed transitions independently of the Zeeman strength. Therefore, twisted light enables selective access to chiral e-e excitation features within the same spin sector while excluding inter-sector transitions, due to its combined angular momentum selection and spin-conserving nature. As demonstrated in Fig.~\ref{fig:calo-coul2}, each resonance can be directly associated with either singlet–singlet or triplet–triplet transitions, depending on the ground-state symmetry determined by the interplay between magnetic field and e-e interaction strength. This clear separation, absent in conventional dipolar excitation, allows TL to resolve singlet–triplet crossings optically and track changes in parity across narrow magnetic field windows. These features make TL spectroscopy a sensitive tool for probing interaction-driven transitions and internal symmetries in strongly correlated quantum systems.

\vskip0.2in
\section{\label{sec:level8}{Conclusions}}

In summary, we have shown how the optical response of the system is governed by the interplay between parity transitions, spin selection rules, and resonance conditions, which together determine the accessibility of angular-momentum–resolved or chiral excitations. Twisted light (TL) hence serves as a selective and sensitive probe of internal excitations and symmetry changes in few-electron quantum systems, capable of resolving spin-sector transitions even within narrow magnetic field windows. More broadly, while the excitation spectrum itself remains unaffected by the Zeeman interaction, the magnetic field intervals where singlet–triplet crossings occur are strongly Zeeman-dependent. The key point is that TL provides a selective and sensitive tool for probing one-photon two-chiral-quanta resonances and detecting parity changes in the excitation spectrum induced by magnetic fields. Beyond resolving such symmetry-driven phenomena, absorption measurements under angular-momentum–structured illumination offer a precise spectroscopic fingerprint of the material’s spin-dependent structure, enabling the characterization of spin textures and magnetic anisotropies through light–matter interaction with orbital angular momentum.

Our detailed calculations show that TL vortex light provides a powerful tool to probe the internal symmetry and interaction effects in two-electron confined systems, for instance in quantum dots under magnetic field effects. Unlike conventional light coupling within the dipolar approximation where the only the CM is responding, TL couples to both CM as well as the relative motion coordinate, thereby revealing interaction-driven dynamics that remain dark otherwise. The observed redshifted resonances, spin-selective absorption, and excitation suppression at high magnetic fields collectively signal the breakdown of Kohn’s theorem and expose the rich internal structure of the system.

We also demonstrated that the analytically solvable Calogero model ($\eta=2$ in an inverse power-law effective e-e interaction) agrees very well with results from the widely employed bare Coulomb-interacting case, reproducing qualitative and quantitatively the same essential physics, thus validating its use as a transparent and versatile theoretical framework. Unlike the Coulomb potential, which generally requires numerical treatment, the Calogero model admits exact analytical solutions for both eigenvalues and eigenfunctions for a two electron system under parabolic confinement. This solvability enables rigorous analysis of optical selection rules, parity transitions, and correlated dynamics, while offering clear physical insight and minimal computational cost. Its close agreement with the Coulomb case further supports its applicability for interpreting experiments and exploring more complex interaction regimes in confined quantum systems.

Taken together, our findings open a path toward exploring symmetry breaking, spin–orbit coupling, strong correlations, and topological phases—such as fractional quantum Hall states—under structured optical driving. Combining beams with distinct or higher-order orbital angular momenta may enable enhanced tunability of the excitation landscape and angular-momentum–driven transitions or generalizations to multi-electron systems and more complex interaction regimes.

\begin{acknowledgments}
F.J.R. and L.Q. acknowledge financial support from the Facultad de Ciencias, Universidad de los Andes, through projects INV-2025-213-3449 and INV-2021-128-2292. F.J.R. also gratefully acknowledges the STAI period (research leave) granted by Universidad de los Andes. 
\end{acknowledgments}
\section{\label{sec:level14}{Appendices}}
We present here a full and detailed analysis of all the mathematics and physics underlying our theoretical results. The range of different approaches,  terminologies, notations and disciplinary backgrounds that we have seen across the attosecond research domain, warrant giving such an integrated and complete analysis to fully support our findings.

\subsection{Appendix A: Oscillator cartesian and chiral ladder operators}

For convenience we recall the definitions introduced in Sec.~\ref{sec:level2}. We start by defining CM-annihilation dimensionless operators $\hat{\vec A}=\left ( \hat{A}_{x}, \hat{A}_{y} \right )$ as:
\begin{eqnarray}
\hat{\vec A}&=&\frac{1}{\sqrt{2}}\left ( \frac{1}{\ell} \hat{\vec R}+\frac{i\ell}{\hbar} \hat{\vec P} \right )
\label{Eq:q5}
\end{eqnarray}
and correspondingly CM-creation operators $\hat{\vec A}^{\dag}$.

Similarly, for the relative motion we define operators $\hat{\vec a}=\left ( \hat{a}_{x}, \hat{a}_{y} \right )$ as:
\begin{eqnarray}
\hat{\vec a}&=&\frac{1}{\sqrt{2}}\left ( \frac{1}{2\ell} \hat{\vec r}+\frac{i2\ell}{\hbar} \hat{\vec p} \right )
\label{Eq:q6}
\end{eqnarray}

The CM-Hamiltonian can be expressed in terms of cartesian ladder operators as:
\begin{eqnarray}
\nonumber \hat{H}_{CM}&=&\hbar \omega \left (  \hat{A}_{x}^{\dag}\hat{A}_{x}+\hat{A}_{y}^{\dag}\hat{A}_{y}+1 \right )\\
&-&\frac{i}{2}\hbar \omega_c \left ( \hat{A}_{x}^{\dag}\hat{A}_{y}-\hat{A}_{x}\hat{A}_{y}^{\dag} \right )\\
\label{Eq:q7}
\end{eqnarray}
Similarly, the relative Hamiltonian turns out to be:
\begin{eqnarray}
\nonumber \hat{H}_{r}&=&\hbar \omega \left (  \hat{a}_{x}^{\dag}\hat{a}_{x}+\hat{a}_{y}^{\dag}\hat{a}_{y}+1 \right )\\
&-&\frac{i}{2}\hbar \omega_c \left ( \hat{a}_{x}^{\dag}\hat{a}_{y}-\hat{a}_{x}\hat{a}_{y}^{\dag} \right )+V_{e-e}(|\hat{\vec{r}}|)\\
\label{Eq:q8}
\end{eqnarray}

Finally, we deploy circular ladder operators to obtain compact expressions for both the CM and relative Hamiltonians. These new operators, \( \hat{A}_{R} \) and \( \hat{A}_{L} \), corresponding to right- and left-circular CM modes, are defined as
\begin{equation}
\hat{A}_{R} = \frac{1}{\sqrt{2}}(\hat{A}_{x} - i\hat{A}_{y}) \; ; \;
\hat{A}_{L} = \frac{1}{\sqrt{2}}(\hat{A}_{x} + i\hat{A}_{y}),
\label{Eq:q9}
\end{equation}
The corresponding Hermitian conjugates \( \hat{A}_{R}^{\dagger} \) and \( \hat{A}_{L}^{\dagger} \) are defined analogously. Similar expressions hold for the relative-coordinate operators:
\begin{equation}
\hat{a}_{R} = \frac{1}{\sqrt{2}}(\hat{a}_{x} - i \hat{a}_{y}) \; ; \;
\hat{a}_{L} = \frac{1}{\sqrt{2}}(\hat{a}_{x} + i \hat{a}_{y})
\label{Eq:q10}
\end{equation}
The CM-Hamiltonian in Eq.(\ref{Eq:q7}) can finally be written as:
\begin{eqnarray}
\hat{H}_{CM}=\hbar \omega +\hbar \omega_R\hat{A}_{R}^{\dag}\hat{A}_{R}+\hbar \omega_L\hat{A}_{L}^{\dag}\hat{A}_{L}
\label{Eq:q11}
\end{eqnarray}
with frequencies $\omega_R=\omega +\frac{\omega_c}{2}$ and $\omega_L=\omega -\frac{\omega_c}{2}$. Hence the CM particle dynamics reduces to that of a two dimensional harmonic oscillator with different chiral right ($\hbar \omega_R$) and left ($\hbar \omega_L$) energy quanta depending on the magnetic field strength.

In a similar way, the relative Hamiltonian in Eq.(\ref{Eq:q8}) transforms to:
\begin{eqnarray}
\hat{H}_{r}=\hbar \omega +\hbar \omega_R\hat{a}_{R}^{\dag}\hat{a}_{R}+\hbar \omega_L\hat{a}_{L}^{\dag}\hat{a}_{L}+V_{e-e}(|\hat{\vec{r}}|)
\label{Eq:q12}
\end{eqnarray}
Thus, due to the presence of the e-e interaction term $V_{e-e}(|\hat{\vec{r}}|)$, the relative particle dynamics corresponds to that of a chirally {\it coupled} 2D oscillator.

From now on, we will ignore the constant term $\hbar \omega$ in the expressions of both CM and relative Hamiltonians in Eqs.(\ref{Eq:q11}-\ref{Eq:q12}).
\subsubsection {Relative motion $su(1,1)$ dynamical symmetry: Chiral ladder operators} 

There is an interesting $su(1,1)$ algebraic dynamical structure of the relative motion of two electrons. Previous attempts to elucidate the ladder operators in this scenario have been limited to 1D systems \cite{agarwal} or simplified versions in 2D situations \cite{dong1}. We start by ignoring the e-e interaction in the relative dynamics. We can construct the relative circular operator combinations:

\begin{align}
\hat{k}_z &= \tfrac{1}{2}(\hat{a}_{R}^{\dag} \hat{a}_{R} + \hat{a}_{L}^{\dag} \hat{a}_{L} + 1) \nonumber \\
\hat{k}^+ &= \hat{a}_{R}^{\dag} \hat{a}_{L}^{\dag} \; ; \;
\hat{k}^- = \hat{a}_{R} \hat{a}_{L}
\label{Eq:q14}
\end{align}

that satisfy the commutation rules:
\begin{eqnarray}
\nonumber \left [\hat{k}_z,\hat{k}^{\pm}\right ]&=&\pm \hat{k}^{\pm}\\
\left [\hat{k}^{+},\hat{k}^{-}\right ]&=&-2 \hat{k}_z
\label{Eq:q15}
\end{eqnarray}
Therefore, the set of operators $\{ \hat{k}_z, \hat{k}^{\pm} \}$ defined in Eq.(\ref{Eq:q14}) yields a two-mode realization of the Lie algebra $su(1,1)$. The Casimir operator is given by:
\begin{eqnarray}
\nonumber \hat{c}&=&\hat{k}_z^2-\frac{1}{2}\left ( \hat{k}^+\hat{k}^-+\hat{k}^-\hat{k}^+ \right )
=
\frac{1}{4} \left ( \left ( \frac{\hat{L}_z}{\hbar} \right )^2-1  \right )
\label{Eq:q15a}
\end{eqnarray}
which is directly connected with a constant of motion, namely the angular momentum $\hat{L}_z$:
\begin{eqnarray}
\frac{\hat{L}_z}{\hbar}=\hat{a}_{R}^{\dag}\hat{a}_{R}-\hat{a}_{L}^{\dag}\hat{a}_{L}\ \ .
\label{Eq:q15b}
\end{eqnarray}
In the presence of the $V_{e-e}(|\hat{\vec{r}}|)= \frac{\tilde {\alpha}}{\hat{r}^2}$ interaction term, a two-mode realization of $su(1,1)$ for the relative motion dynamical symmetry is still possible. Define the following operators in terms of dimensionless interaction strength $\alpha=\frac{\tilde {\alpha}/\ell_o^2}{\hbar\omega_o}$ and operator $\hat{\bar r}^2$ (see Eq.(\ref{Eq:q13})):
\begin{eqnarray}
\nonumber \hat{K}_z&=&\hat{k}_z+\frac{\alpha}{4\hat{\bar r}^2}\\
\hat{K}^{\pm}&=& \hat{k}^{\pm}-\frac{\alpha}{4\hat{\bar r}^2}
\label{Eq:q16}
\end{eqnarray}
It can easily be verified that these operators satisfy the $su(1,1)$ Lie algebra:
\begin{eqnarray}
\nonumber \left [\hat{K}_z,\hat{K}^{\pm}\right ]&=&\pm \hat{K}^{\pm}\\
\left [\hat{K}^{+},\hat{K}^{-}\right ]&=&-2 \hat{K}_z
\label{Eq:q17}
\end{eqnarray}
Now, the Casimir oparator is:
\begin{eqnarray}
\nonumber \hat{\cal {C}}&=&\hat{K}_z^2-\frac{1}{2}\left ( \hat{K}^+\hat{K}^-+\hat{K}^-\hat{K}^+ \right )\\
&=&\hat{c}+\frac{\alpha}{4}
\label{Eq:q17a}
\end{eqnarray}
where $\hat{c}$ is given in Eq.(\ref{Eq:q15a}).
Since $V_{e-e}(|\hat{\vec{r}}|)= \frac{\tilde{\alpha}}{\hat{r}^2}=2\hbar\omega_0\left (\frac{\alpha}{4\hat{\bar r}^2}\right )$, the relative Hamiltonian in Eq.(\ref{Eq:q12}) can finally be written as:

\begin{equation}
\hat{H}_{r}=\hbar \omega \left ( 2\hat{K}_z-1 \right )+\frac{\omega_c}{2} \hat{l}_z
\label{Eq:q19}
\end{equation}
Therefore, the dynamical $su(1,1)$ symmetry of the relative Hamiltonian is 
established.
\vskip0.1in

\subsection{Appendix B: Interacting eigenstates in terms of non-interacting ones}

In the case of a non-interacting two electron QD, i.e. $\alpha=0$, the relative Hamiltonian in Eq.(\ref{Eq:q12}) is immediately diagonalized by a state basis formed by number 
eigenstates for each one 
of the right and left oscillators. Thus non-interacting eigenstates can be expressed as $|n_R,n_L\rangle$ that are angular momentum eigenstates:

\begin{eqnarray}
\nonumber \hat{H}_{r}
|n_R,n_L\rangle&=&
\hbar \omega_0\left (n_R +n_L\right )|n_R,n_L\rangle\\
\hat{L}_z|n_R,n_L\rangle&=&(n_R-n_L)\hbar |n_R,n_L\rangle
\label{Eq:q32}
\end{eqnarray}

with $n_R,n_L=0,1,2,...$. From Eq.(\ref{Eq:q15b}),  $m=n_R-n_L$.
We can express interacting states $|n,m\rangle_{\alpha}$ in terms of non-interacting ones $|n_R,n_L\rangle$ as follows:

\begin{eqnarray}
|0,m\rangle_{\alpha}= \sum_{n=0}^{\infty}C_n|n,n+|m|\rangle
\label{Eq:q33}
\end{eqnarray}

In order to get the coefficients $C_n$,  insert the latter expression into Eq.(\ref{Eq:q29}) to obtain a three-term recursion relation:
\begin{widetext}
\begin{eqnarray}
\nonumber C_n \left (n(n+|m|) -\frac{\alpha}{4}\right )+C_{n+1}\left ( 2n+|m|+1 \right )\sqrt{(n+1)(n+|m|+1)}+C_{n+2}\sqrt{(n+1)(n+2)(n+|m|+1)(n+|m|+2)}=0\\
\label{Eq:q34}
\end{eqnarray}
\end{widetext}
The solution $C_n$ to the above recursion equation is:
\begin{eqnarray}
C_n=\sqrt{\frac{n!}{(n+|m|)!}}\,\,\binom{\frac{\alpha_m+m}{2}}{n}
\label{Eq:q35}
\end{eqnarray}
where $\alpha_m=\sqrt{m^2+\alpha}$ and $\binom{\nu}{n}$ the binomial coefficient.
The normalized lowest energy state for a given $m\le 0$-sector, for example, can be expressed as:
\begin{widetext}
\begin{align}
|0,m\rangle_{\alpha} &= 
\frac{\Gamma\left ( \frac{\alpha_m+|m|}{2}+1 \right )}{\sqrt{\Gamma \left ( \alpha_m+1 \right )}} 
\sum_{n=0}^{\infty}\sqrt{\frac{n!}{(n+|m|)!}} \binom{\frac{\alpha_m-|m|}{2}}{n} |n,n+|m|\rangle \nonumber \\
&= 
\frac{\Gamma\left ( \frac{\alpha_m-|m|}{2}+1 \right )}{\sqrt{\Gamma \left ( \alpha_m+1 \right )}} 
L_{\frac{\alpha_m-|m|}{2}}^{|m|}\left ( -\hat{k}^{+} \right ) \hat{a}_{L}^{\dag\,|m|} |0,0\rangle
\label{Eq:sq36}
\end{align}
\end{widetext}

where $L_{\nu}^{-m}(x)$ is a Laguerre function. The energy $E_{0,m}(\alpha)$ turns out to be:

\begin{eqnarray} 
E_{0,m}(\alpha)= \,_{\alpha} \langle0,m| \hat{H}_{r}|0,m  \rangle_{\alpha}
= \hbar \omega \alpha_m+m \frac{\hbar \omega_c}{2}
\label{Eq:sq37}
\end{eqnarray}
The mean number of chiral excitations is given by $\,_{\alpha} \langle 0,m| \hat{a}_{R}^{\dag}\hat{a}_{R}|0,m  \rangle_{\alpha}$ and $\,_{\alpha} \langle 0,m| \hat{a}_{L}^{\dag}\hat{a}_{L}|0,m  \rangle_{\alpha}$ where:
\begin{eqnarray}
\nonumber \,_{\alpha} \langle 0,m| \hat{a}_{R}^{\dag}\hat{a}_{R}|0,m  \rangle_{\alpha}&=&\frac{\left ( \alpha_m-|m| \right )^2}{4\alpha_m}\\
\,_{\alpha} \langle 0,m| \hat{a}_{L}^{\dag}\hat{a}_{L}|0,m  \rangle_{\alpha}&=&\frac{\left (\alpha_m+|m| \right )^2}{4\alpha_m}
\label{Eq:q38}
\end{eqnarray}
where $ \langle \hat{a}_{R}^{\dag}\hat{a}_{R}  \rangle-\langle \hat{a}_{L}^{\dag}\hat{a}_{L}  \rangle=-|m|$ as expected. It is straightforward to get closed expressions for excited states in a given $m$-sector by applying the $su(1,1)$ ladder operator $\hat{K}^+$
which leads to normalized states ($n \ge 0$):
\begin{eqnarray}
|n,m  \rangle_{\alpha}=\sqrt{\frac{\Gamma \left ( \alpha_m+1  \right )}{\Gamma \left (\alpha_m+n+1  \right )}}\frac{\hat{K}^{+n}}{\sqrt{n!}}|0,m  \rangle_{\alpha}\ \ .
\label{Eq:sq1}
\end{eqnarray}

\subsection{Appendix C: Validation of $su(1,1)$ Lie algebra structure}

We now prove that the relative motion operators displayed in Eq.(\ref{Eq:q16}) for a $V_{e-e}(|\hat{\vec{r}}|)= \frac{\tilde{\alpha}}{\hat{r}^2}=2\hbar\omega_0\left (\frac{\alpha}{4\hat{\bar r}^2}\right )$ interaction, indeed fulfill the $su(1,1)$ Lie algebra requirements.\\
(i) Calculation of the commutator $\left [\hat{K}^{+},\hat{K}^{-}\right ]$:
\begin{eqnarray}
\left [\hat{K}^{+},\hat{K}^{-}\right ]=-2\hat{k}_z-\frac{\alpha}{4}\left [\hat{k}^{+}-\hat{k}^{-},\frac{1}{\hat{\bar r}^2}\right ]
\label{Eq:q20}
\end{eqnarray}
Using the identity:
\begin{equation}
\frac{1}{\hat{\bar r}^2}=\int_0^{\infty}e^{-s\hat{\bar r}^2}ds
=\int_0^{\infty}ds\sum_{j=0}^{\infty}\frac{(-s\hat{\bar r}^2)^j}{j!}
\label{Eq:q21}
\end{equation}
yields 
\begin{eqnarray}
\left [\hat{k}^{+}-\hat{k}^{-},\hat{\bar r}^{2j}\right ]=-2j\hat{\bar r}^{2j}
\label{Eq:q22}
\end{eqnarray}
and the commutator on the right hand side of Eq.(\ref{Eq:q20}) turns out to be:
\begin{eqnarray}
\left [\hat{k}^{+}-\hat{k}^{-},\frac{1}{\hat{\bar r}^2}\right ]=\frac{2}{\hat{\bar r}^2}
\label{Eq:q23}
\end{eqnarray}
Consequently:
\begin{eqnarray}
\left [\hat{K}^{+},\hat{K}^{-}\right ]=-2\hat{K}_z
\label{Eq:q24}
\end{eqnarray}
as expected.\\
(ii) Calculation of the commutator $\left [\hat{K}_z,\hat{K}^{\pm}\right ]$:
\begin{eqnarray}
\left [\hat{K}_z,\hat{K}^{\pm}\right ]=\pm \hat{k}^{\pm}-\frac{\alpha}{4}\left [\hat{k}_z+\hat{k}^{\pm},\frac{1}{\hat{\bar r}^2}\right ]
\label{Eq:q25}
\end{eqnarray}
Procceding in a similar way as above we find:
\begin{eqnarray}
\left [\hat{k}_z+\hat{k}^{\pm},\frac{1}{\hat{\bar r}^2}\right ]=\pm \frac{1}{\hat{\bar r}^2}
\label{Eq:q26}
\end{eqnarray}
which leads to
\begin{eqnarray}
\left [\hat{K}_z,\hat{K}^{\pm}\right ]=\pm \hat{K}^{\pm}\ \ .
\label{Eq:q27}
\end{eqnarray}
Hence the set of operators $\{ \hat{K}_z,\hat{K}^{\pm} \}$ constitutes a valid realization of the $su(1,1)$ Lie algebra. These new generators are therefore useful for analyzing the relative motion of two electrons in presence of an interaction of the form $V_{e-e}(|\hat{\vec{r}}|)=2\hbar\omega_0\left (\frac{\alpha}{4\hat{\bar r}^2}\right )$ as is evident in Eq.(\ref{Eq:q19}).\\

\subsection{Appendix D: Action of $su(1,1)$ generators on relative Hamiltonian eigenstates $|n,m\rangle_{\alpha}$}
Stationary states $|n,m\rangle_{\alpha}$ are eigenstates of the diagonal generator $\hat{K}_z$ too:
\begin{eqnarray}
\hat{K}_z|n,m\rangle_{\alpha}=\frac{1}{2}\left ( 2n+\alpha_m+1 \right )|n,m\rangle_{\alpha}
\label{Eq:b10}
\end{eqnarray}
where $\alpha_m=\sqrt{m^2+\alpha}$.
From Eqs.(\ref{Eq:q15a}) and (\ref{Eq:q17a}) it can easily be verified that :
\begin{eqnarray}
\nonumber \hat{K}^{+}\hat{K}^{-}&=&\hat{K}_z^2-\hat{K}_z-\frac{1}{4}\left (\alpha_m^2-1 \right )\\
\hat{K}^{-}\hat{K}^{+}&=&\hat{K}_z^2+\hat{K}_z-\frac{1}{4}\left ( \alpha_m^2-1 \right )
\label{Eq:b11}
\end{eqnarray}
which allows to prove that $\hat{K}^{\pm}$ act as ladder operators:
\begin{eqnarray}
\nonumber \hat{K}^{+}|n,m\rangle_{\alpha}&=&\sqrt{(n+1)(n+1+\alpha_m)}\,|n+1,m\rangle_{\alpha}\\
\hat{K}^{-}|n,m\rangle_{\alpha}&=&\sqrt{n(n+\alpha_m)}\,|n-1,m\rangle_{\alpha}
\label{Eq:b12}
\end{eqnarray}
For a given $m$-sector ($m \le 0$), the reiterated application of the first line of Eqs.(\ref{Eq:b12}) leads to a closed expression for any normalized eigenstate $|n,m\rangle_{\alpha}$ in terms of the lowest energy eigenstate $|0,m\rangle_{\alpha}$, Eq.(\ref{Eq:sq36}):
\begin{eqnarray}
|n,m\rangle_{\alpha}&=&\sqrt{\frac{\Gamma(\alpha_m+1)}{\Gamma(\alpha_m+n+1)}}\, \frac{\hat{K}^{+ n}}{\sqrt{n!}}|0,m\rangle_{\alpha}
\label{Eq:b13}
\end{eqnarray}
Hence the expected value, in any of the eigenstates $|n,m\rangle_{\alpha}$, of the relative e-e separation squared is:
\begin{eqnarray}
\nonumber \langle \hat{r}^2  \rangle&=&4\ell_0^2\,_{\alpha} \langle n,m| \hat{\bar r}^2 |n,m\rangle_{\alpha}\\
\nonumber &=&4\ell_0^2\,_{\alpha} \langle n,m| 2\hat{K}_z+\hat{K}^++\hat{K}^- |n,m\rangle_{\alpha}\\
\nonumber &=&4\ell_0^2\,_{\alpha} \langle n,m|2\hat{K}_z |n,m\rangle_{\alpha}\\
&=&4\ell_0^2\left ( 2n+\alpha_m+1 \right )
\label{Eq:b14}
\end{eqnarray}
The spatial 'size' of an eigenstate $|n,m\rangle_{\alpha}$ can be estimated in a certain sense by the square root of the expectation value of $\langle \hat{\vec{r}}_1^{\, 2}  \rangle=\langle \hat{\vec{r}}_2^{\, 2}  \rangle$. Thus,
\begin{eqnarray}
\nonumber \langle \hat{\vec{r}}_1^{\, 2}  \rangle&=&\,_{\alpha} \langle n,m| \hat{R}^2 +\frac{\hat{r}^2}{4}+\vec{R}\cdot \vec{r}|n,m\rangle_{\alpha}\\
\nonumber &=&\langle CM_{GS}| \hat{R}^2 |CM_{GS}\rangle+\,_{\alpha} \langle n,m| \frac{\hat{r}^2}{4}|n,m\rangle_{\alpha}\\
&=&\ell_0^2\left ( 2n+\alpha_m+2 \right )
\label{Eq:b15}
\end{eqnarray}
In reaching the last result we have assumed the center-of-mass stays in its ground state, i.e. $\langle CM_{GS}| \hat{R}^2 |CM_{GS}\rangle=\ell_0^2$.
Since $\hat{r}^2=\hat{r}_1^{\, 2}+\hat{r}_2^{\, 2}-2\hat{\vec{r}}_1\cdot \hat{\vec{r}}_2$, the angular e-e correlation can be estimated from:
\begin{eqnarray}
\nonumber \frac{\langle \hat{\vec{r}}_1\cdot \hat{\vec{r}}_2\rangle}{\langle \hat{\vec{r}}_1^{\, 2}  \rangle}&=&1-\frac{1}{2}\frac{\langle \hat{\vec{r}}^{\, 2}  \rangle}{\langle \hat{\vec{r}}_1^{\, 2}  \rangle}\\
&=&-\frac{2n+\alpha_m }{2n+\alpha_m+2}
\label{Eq:b16}
\end{eqnarray}
The negative sign is reasonable, since the e-e repulsion enforces the mutual electron angle to become close to $\pi$ radians.
Eqs.(\ref{Eq:b14})-(\ref{Eq:b16}) coincide with results previously reported in \cite{qaj} where a wave function approach was used.

For the sake of completeness, we now show expressions for positive relative angular momentum. In this case $n_{min}=m$. Thus:
\begin{eqnarray}
|m,m\rangle_{\alpha}= \sum_{n=0}^{\infty}C_n|n+m,n\rangle
\label{Eq:a33}
\end{eqnarray}
The recursion relation Eq.(\ref{Eq:q34})
allows a closed form for $C_n$ as:
\begin{eqnarray}
C_n=\sqrt{\frac{(n+m)!}{n!}}\,\,\binom{\frac{\alpha_m+m}{2}}{n+m}
\label{Eq:a35}
\end{eqnarray}
Normalized lowest energy states, for a given $m$-tower, in terms of non-interacting $\{ |n_R,n_L\rangle \}$ states become:
\begin{eqnarray}
\nonumber |m,m\rangle_{\alpha}&=&\frac{\Gamma\left ( \frac{\alpha_m-m}{2}+1 \right )}{\sqrt{\Gamma \left (\alpha_m+1 \right )}} \sum_{n=0}^{\infty}\sqrt{\frac{(n+m)!}{n!}}\,\,\binom{\frac{\alpha_m+m}{2}}{n+m}|n+m,n\rangle\\
&=&\sqrt{m!}\frac{\Gamma\left ( \frac{\alpha_m-m}{2}+1 \right )}{\sqrt{\Gamma \left (\alpha_m+1 \right )}}L_{\frac{\alpha_m-m}{2}}^{m}\left ( -\hat{k}^{+}\right )|m,0\rangle
\label{Eq:a36}
\end{eqnarray}

\subsection{Appendix E: Real space representation}
For a real space representation of the quantum dot $su(1,1)$ algebra, polar coordinates $(r,\phi)$ for the relative particle will be used.
For non-interacting electrons ($\alpha=0$), the normalized real space wavefunctions are:
\begin{widetext}
\begin{eqnarray}
\nonumber \psi_{n,n+|m|}(r,\phi)&=&\langle r,\phi|n,n+|m|\rangle
=(-1)^n\frac{e^{im\phi}}{2\sqrt{\pi}\ell_0}\sqrt{\frac{n!}{(n+|m|)!}}\left ( \frac{r}{2\ell_0}\right )^{|m|}e^{-\frac{r^2}{8\ell_0^2}}L_n^{|m|}\left ( \frac{r^2}{4\ell_0^2} \right )
\label{Eq:a41}
\end{eqnarray}
In order to find the corresponding wavefunctions for interacting electrons ($\alpha>0$) a useful identity is:
\begin{eqnarray}
x^{\alpha_m}=x^{|m|}\Gamma\left ( \frac{\alpha_m+|m|}{2}+1 \right )
\sum_{n=0}^{\infty}(-1)^n\frac{n!}{(n+|m|)!}\,\,\binom{\frac{\alpha_m-|m|}{2}}{n}L_n^{|m|}\left ( x^2 \right )
\label{Eq:a42}
\end{eqnarray}
\end{widetext}
where $\alpha_m=\sqrt{\alpha+m^2}$. The real space wavefunction corresponding to the lowest energy eigenstate of the $m$-sector given by Eq.(\ref{Eq:sq36}), and using Eqs.(\ref{Eq:a41})-(\ref{Eq:a42}), is found to be:
\begin{widetext}
\begin{eqnarray}
\nonumber \psi_{0,m}^{(\alpha)}(r,\phi)&=&\langle r,\phi|0,m\rangle_{\alpha} \\
\nonumber &=&\frac{e^{im\phi}}{2\sqrt{\pi}\ell_0}\left ( \frac{r}{2\ell_0}\right )^{|m|}e^{-\frac{r^2}{8\ell_0^2}}\frac{\Gamma\left ( \frac{\alpha_m+|m|}{2}+1 \right )}{\sqrt{\Gamma \left ( \alpha_m+1 \right )}} \sum_{n=0}^{\infty}(-1)^n\frac{n!}{(n+|m|)!}\,\,\binom{\frac{\alpha_m-|m|}{2}}{n}\,L_n^{|m|}\left ( \frac{r^2}{4\ell_0^2} \right )\\
&=&\frac{e^{im\phi}}{2\sqrt{\pi}\ell_0}\frac{1}{\sqrt{\Gamma \left ( \alpha_m+1 \right )}}\left ( \frac{r}{2\ell_0}\right )^{\alpha_m}e^{-\frac{r^2}{8\ell_0^2}}
\label{Eq:a43}
\end{eqnarray}
\end{widetext}
According to Eqs.(\ref{Eq:q6}) and (\ref{Eq:q10}):
\begin{eqnarray}
\nonumber \hat{a}_{R}&\rightarrow&\frac{e^{-i\phi}}{2}\left (\frac{r}{2\ell_0}+2\ell_0\frac{\partial}{\partial\, r}-i\frac{2\ell_0}{r}\frac{\partial}{\partial\, \phi}\right )\\
\hat{a}_{L}&\rightarrow&\frac{e^{i\phi}}{2}\left (\frac{r}{2\ell_0}+2\ell_0\frac{\partial}{\partial\, r}+i\frac{2\ell_0}{r}\frac{\partial}{\partial\, \phi}\right )
\label{Eq:a37}
\end{eqnarray}
yielding 
\begin{eqnarray}
\hat{k}^+-\hat{k}^-&\rightarrow&-\left ( r\frac{\partial}{\partial\, r}+1 \right )
\label{Eq:a38}
\end{eqnarray}
A straightforward calculation leads to
\begin{eqnarray}
\nonumber \hat{K}^+&=& \hat{k}^+-\frac{\alpha}{4\hat{\bar r}^2}\\
\nonumber &=& \hat{k}^++ \hat{k}_z-\hat{K}_z\\
&=&\frac{1}{2}\left (2\hat{k}^+- \hat{k}^-+ \hat{k}^-+ 2\hat{k}_z-2\hat{K}_z  \right )
\label{Eq:a39}
\end{eqnarray}
and consequently to:
\begin{eqnarray}
\nonumber \hat{K}^+|n,m\rangle_{\alpha}=\frac{1}{2}\left (\hat{k}^+- \hat{k}^-+ \hat{\bar r}^2-\left ( 2n+\alpha_m+1 \right )  \right )|n,m\rangle_{\alpha} \ \ .
\label{Eq:a45}
\end{eqnarray}
Therefore, the real space representation of $\hat{k}^+-\hat{k}^-$ given in Eq.(\ref{Eq:a38}) allows us to write:
\begin{eqnarray}
\nonumber \hat{K}^+\psi_{n,m}^{(\alpha)}(r,\phi)=-\frac{1}{2}\left ( r\frac{\partial}{\partial\, r}-\frac{r^2}{4\ell_0^2}+2(n+1)+\alpha_m \right )\psi_{n,m}^{(\alpha)}(r,\phi)\\
\label{Eq:a40}
\end{eqnarray}
with $\psi_{n,m}^{(\alpha)}(r,\phi)=\langle r,\phi|n,m\rangle_{\alpha}$.
Finally, by applying the raising ladder operator $\hat{K}^+$ as given by Eq.(\ref{Eq:a40}), the whole real space wavefunction spectrum for a given $m$-sector can be found:
\begin{widetext}
\begin{eqnarray}
\nonumber \psi_{n,m}^{(\alpha)}(r,\phi)&=&\sqrt{\frac{\Gamma(\alpha_m+1)}{\Gamma(\alpha_m+n+1)}}\, \langle r,\phi|\frac{\hat{K}^{+ n}}{\sqrt{n!}}|0,m\rangle_{\alpha} \\
\nonumber &=&\frac{e^{im\phi}}{2\sqrt{\pi}\ell_0}\sqrt{\frac{n!}{\Gamma \left ( \alpha_m+n+1 \right )}}\frac{(-1)^n}{2^nn!}\prod_{j=1}^n\left (  r\frac{\partial}{\partial\, r}-\frac{r^2}{4\ell_0^2}+2j+\alpha_m \right )\left [ \left ( \frac{r}{2\ell_0}\right )^{\alpha_m}e^{-\frac{r^2}{8\ell_0^2}}\right ]\\
&=&(-1)^n\frac{e^{im\phi}}{2\sqrt{\pi}\ell_0}\sqrt{\frac{n!}{\Gamma \left ( \alpha_m+n+1 \right )}}\,\left ( \frac{r}{2\ell_0}\right )^{\alpha_m}e^{-\frac{r^2}{8\ell^2}}\,L_n^{\alpha_m}\left (\frac{r^2}{4\ell_0^2} \right )\ \ .
\label{Eq:a44}
\end{eqnarray}
\end{widetext}

\subsection{Appendix F: Comparison between Calogero and Coulomb interactions}
\label{appf}
To benchmark the reliability of the analytically solvable model, we compare its predictions against the full Coulomb interaction case. Despite their distinct forms, both interaction potentials yield similar qualitative and quantitative features in the energy spectrum, particularly under confinement and magnetic field effects.

\begin{figure}[hbt]
    \centering
        \includegraphics[width=0.5\textwidth]
{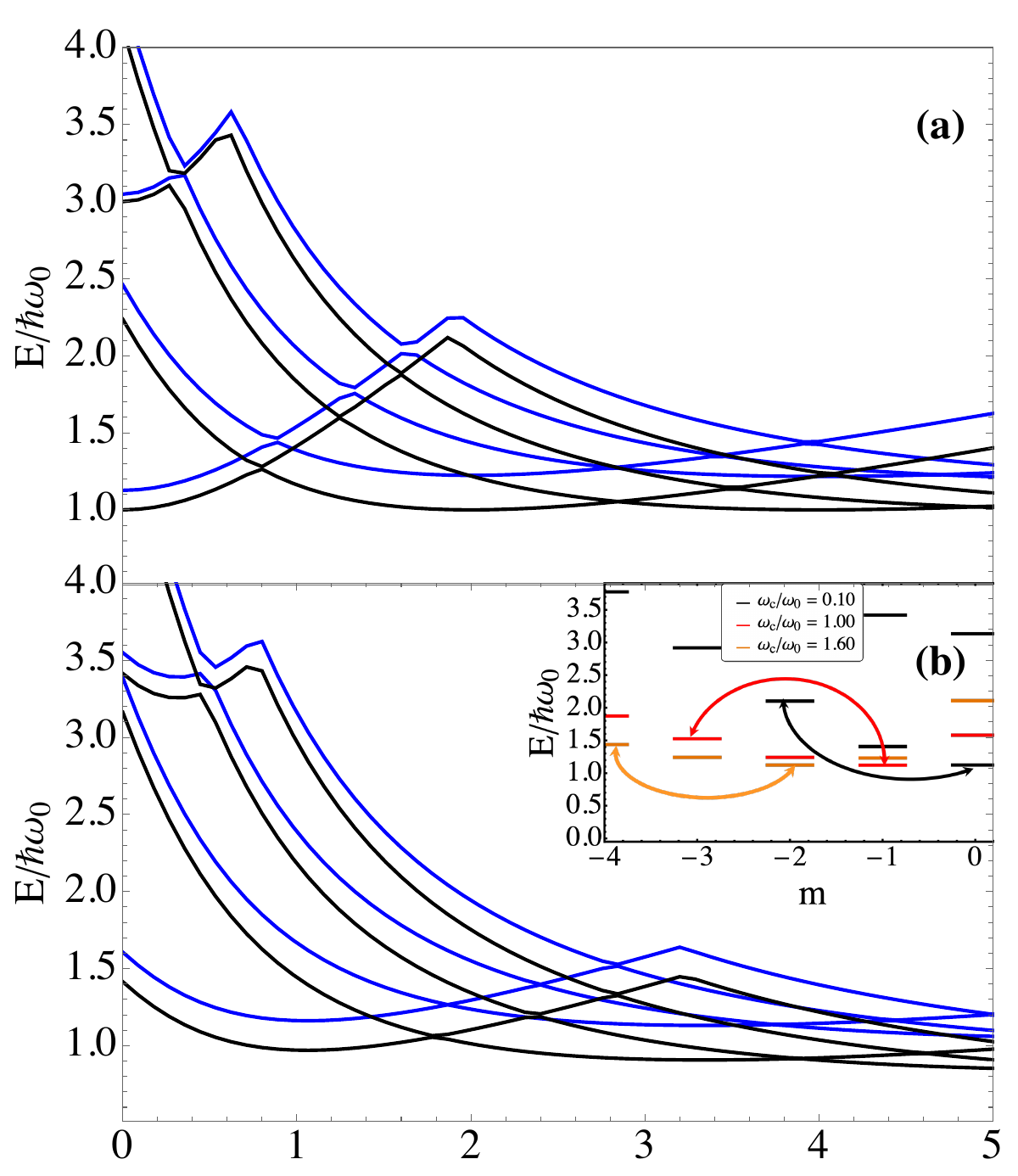}

    \caption{Relative coordinates energy spectrum \( E_n(\alpha, m)/\hbar\omega_0 \) as a function of the normalized cyclotron frequency \( \omega_c/\omega_0 \), for two types of e-e interactions (\( \sim 1/r^\eta )\):  Coulomb potential (\( \eta =1 \), blue curves) and  Calogero potential (\( \eta=2 \), black curves). The interaction strength is fixed at \( \alpha = 1.25 \). (a) even-parity states (spin singlets), (b) odd-parity states (spin triplets).Inset: Energy levels \( E/\hbar\omega_0 \) as a function of the angular momentum quantum number \( m \) for a two-electron quantum dot with inverse-square interaction. The results are shown for three values of the dimensionless magnetic field: \( \omega_c/\omega_0 = 0.20 \) (black), \( 1.00 \) (red), and \( 1.56 \) (orange). The arrows highlight representative interlevel transitions that conserve \( m \) and connect states with different \( n \) at distinct magnetic field strengths. These transitions illustrate the evolution of the level structure as \( \omega_c \) increases and identify relevant crossings associated with magneto-optical excitations.
}
\label{fig:fig2} 
\end{figure}

Figure~\ref{fig:fig2} compares the energy spectrum obtained using the analytically solvable Calogero interaction (black curves) with the numerically computed Coulomb case (blue curves). Fig~\ref{fig:fig2}~(a) shows the spin singlet spectrum, while Fig~\ref{fig:fig2}~(b) displays the corresponding spin triplet spectrum, both plotted as a function of the normalized magnetic field \( \omega_c/\omega_0 \). The inset in panel~(b) illustrates the energy levels as a function of the relative angular momentum \( m \) for several representative magnetic field values, revealing the magnetic-field–dependent evolution of the low-energy spectrum, for a dimensionless interaction strength ($\alpha=1.25$). The results show that both interaction models: Coulomb ($( (1/r)$  and Calogero-type $(1/r^2)$) exhibit qualitatively similar behaviors across a wide range of magnetic field strengths confirming that the Calogero model captures the essential physics of the interacting QD system while allowing for a full analytical tractability. Although the Coulomb interaction produces a global upward shift in energy levels due to its long-range nature, the ordering of levels, spectral crossings, and overall magnetic-field dependence remain essentially unchanged for the two analyzed e-e potential cases.

As the magnetic field increases, the energy levels in both models tend to flatten or rise slowly, indicating that the cyclotron energy becomes the dominant contribution to the spectrum.
However, the \textit{Zeeman interaction}, which becomes more significant at high magnetic fields, lowers the energy of  triplet states by \( -g^* \mu_B B \), while leaving singlet states unaffected. This is particularly relevant in Fig.~\ref{fig:fig2} (b), where the \( m = 1 \) states are generally associated with  triplets. The increasing contribution of the Zeeman term energetically favors these states, potentially allowing them to cross below the singlet ground state at sufficiently high fields. Unlike in the \( m = 1 \) case, the \( m = 0 \) 
singlet states in Fig.~\ref{fig:fig2}(a) remain unaffected by Zeeman splitting. As a result, at high magnetic fields, the Zeeman effect lowers the energy of triplet states, allowing them to become the ground state. This can lead to a reordering of the spectrum and modify the symmetry and optical properties of the lowest-energy configuration.

\bibliography{twistedlightVF25082025.bib}
\end{document}